\documentclass[pre,aps,superscriptaddress,showpacs,floatfix]{revtex4}
\usepackage{amsmath}
\usepackage{amssymb}
\usepackage{amsbsy}
\usepackage{graphicx}
\usepackage{color}

\usepackage{enumerate}
\usepackage{mathtools}

\usepackage{natbib}

\newcommand{\erfc}{\mathrm{erfc}\,}

\def\be{\begin{equation}}
\def\ee{\end{equation}}
\def\bfi{\begin{figure}}
\def\efi{\end{figure}}
\def\bea{\begin{eqnarray}}
\def\eea{\end{eqnarray}}

\begin{document}

\title{Relation between Statics and Dynamics in the Quench of the Ising Model to below the Critical Point}  

\author{Annalisa Fierro}
\email{annalisa.fierro@spin.cnr.it}
\affiliation{CNR-SPIN, c/o Complesso di Monte S. Angelo, via Cinthia - 80126 - Napoli, Italy}

\author{Antonio Coniglio} 
\email{antonio.coniglio@gmail.com}
\affiliation{CNR-SPIN, c/o Complesso di Monte S. Angelo, via Cinthia - 80126 - Napoli, Italy}

\author{Marco Zannetti}
\email{mrc.zannetti@gmail.com}
\affiliation{Dipartimento di Fisica "E. R. Caianiello", 
Universit\`a di Salerno, Via Giovanni Paolo II 132, I-84084 Fisciano (SA), Italy}

\begin{abstract}

The standard phase-ordering process is obtained by quenching a system, 
like the Ising model, to below the critical point. This is usually done
with periodic boundary conditions to insure ergodicity breaking 
in the low temperature phase.
With this arrangement the infinite system
is known to remain permanently out of equilibrium, i.e.
there exists a well defined asymptotic state which is time-invariant
but different from the ordered ferromagnetic state. 
In this paper we establish the critical nature
of this invariant state, by demonstrating numerically that the quench dynamics 
with periodic and antiperiodic boundary conditions  are indistinguishable one from the other.
However while the asymptotic state does not coincide with the equilibrium state for the periodic case, 
it coincides instead with the equilibrium state of the antiperiodic case, which in fact is critical.   
The specific example of the Ising model is shown to be one instance of a more general
phenomenon, since an analogous picture emerges in the
spherical model, where boundary conditions are kept fixed to periodic, while
the breaking or preserving of ergodicity is managed by imposing the spherical
constraint either sharply or smoothly.


\end{abstract} 

\maketitle

\section{Introduction}

Macroscopic systems, in absence of an external drive, equilibrate with
the environment.
However, relaxation may be slow,
i.e. with a relaxation time which exceeds any attainable observation time~\cite{Palmer}. 
In that case,
only dynamical properties are accessible to observation
and the question naturally arises of what can
be learnt about equilibrium from dynamics. Paradigmatical examples of slow relaxation
are glassy systems~\cite{BCKM} or systems undergoing 
phase ordering after a sudden temperature quench from above 
to below the critical point~\cite{BCKM,Bray}. 
Here we shall look at the problem in the latter
context, whose prototypical instance is the quench of a
ferromagnetic system. In order to make the
presentation as simple as possible, we shall mostly concentrate on
the Ising model. The extension to other phase-ordering systems will be 
discussed at the end of the paper, with the example of the spherical model.

Phase ordering in the Ising model by now is a mature subject, generally considered
to be well understood. For reviews see Refs.~\cite{Bray,Puri,Jo,Henkel}.
Among the many interesting features of the process, in this paper we 
shall be primarily concerned with the lack of equilibration in any finite time,
if the system is infinite. This is frequently referred
to with the catchy expression that the system remains permanently out of equilibrium,
whose meaning, however, has never been fully clarified.
For instance, a similar circumstance arises also when the quench is made to the 
critical temperature $T_c$, because, due to critical slowing down, 
again equilibrium is not reached in any
finite time. Nonetheless, in that case, the process cannot be regarded as 
substantially different from one of equilibration,
because as time grows the system gets 
closer and closer to the equilibrium critical state, which is unique in the sense that
in the thermodynamic limit it is independent of the boundary conditions (BC).
Instead, in the quench to below $T_c$ the picture is qualitatively different,
because, although the state extrapolated from dynamics is unique, the same cannot 
be said of the equilibrium state, which
depends on BC even in the thermodynamic limit. This we have shown
in Ref.~\cite{FCZ} (to be referred
to as I in the following), where we have investigated 
the nature of the equilibrium state in the Ising model below 
$T_c$, under different symmetry-preserving BC.
We have found that while periodic boundary conditions (PBC)
lead to the usual ferromagnetic ordering, due to
the breaking of ergodicity with the consequential spontaneous breaking
of the $\mathbb{Z}_2$ up-down symmetry, 
the scenario changes dramatically with antiperiodic boundary conditions
(APBC), because
ergodicity breaking is precluded. Then,
the system cannot order and complies with the requirement of the transition
by remaining critical also below $T_c$,  all the way down to $T=0$.
We have argued that this new transition, without spontaneous symmetry breaking
and without ordering,
consists in the condensation of fluctuations. In the $1d$ case, since $T_c=0$, the low temperature 
phase is shrunk to just $T=0$.

Motivated by the existence of such diversity in the equilibrium properties, 
in this paper we address
the next natural question, formulated in the title of the paper, of matching
statics and dynamics.
Using the equal-time correlation function as the probing observable, we shall see
that the asymptotic state, extrapolated 
from dynamics, that is by taking the
$t \to \infty$ limit after the thermodynamic limit, is unique and {\it critical}. 
Now, the point is that this, which we may call the time-asymptotic state
and which, we emphasize, is the same for both choices of BC,
is found to coincide with the bona fide equilibrium state, i.e. the one
computed from equilibrium statistical mechanics, in the APBC case but
to be remote from it in the PBC case.
Thus, we have the one and the same 
dynamical evolution which, although not reaching equilibrium in any finite time,
turns out to be informative of the true equilibrium state in one case
(APBC), but not in the other (PBC).
It is, then, appropriate to regard the APBC case as one in which 
equilibrium is approached,
just as in a quench to $T_c$, while the PBC case offers an instance of a system remaining
permanently out of equilibrium. The poor performance in approaching equilibrium
with PBC is traceable to the presence
of ergodicity breaking at the working temperature, which, instead, is preserved when 
APBC are applied.
At the end of the paper we shall argue that the connection
between the presence/absence of ergodicity breaking and the absence/presence of equilibration goes beyond the Ising example, by showing that it takes place 
with the same features in the rather different context of
the spherical and mean-spherical model.

The paper is organized as follows: in section~\ref{II} we formulate the problem. 
In section~\ref{III}  the relation between equilibrium and relaxation in the quench
to above $T_c$ is analyzed by using scaling
arguments. The cases of the quench to $T_c$ with $d=2$, to $T=0$ with $d=1$ 
and to below $T_c$ with $d=2$ are analyzed in sections~\ref{IV}, \ref{V} and
\ref{VI}, respectively. The spherical and mean spherical model are introduced and
investigated in section~\ref{SM}. Concluding remarks are made in section~\ref{CR}.

\section{The problem}
\label{II}

We are concerned with the relaxation dynamics of a system initially prepared in
an equilibrium state at the temperature $T_I$ and suddenly quenched to the 
lower temperature $T_F$. We consider 
the Ising model on a lattice of size $V=L^d$, with the usual nearest neighbours 
interaction
\be
{\cal H}(\boldsymbol{s}) = -J\sum_{<ij>} s_is_j,
\label{Ham.1}
\ee
where $J > 0$ is the ferromagnetic coupling,
$\boldsymbol{s} = [s_i]$ is a configuration of spin variables $s_i=\pm1$
and $<ij>$ is a pair of nearest neighbours. 
We shall study the $d=1$ and $d=2$ cases, where
in the thermodynamic limit there is a critical point at $T_c=0$ and
$T_c=2.269J$, respectively. 
Since the system's size is finite, BC must be specified and, because of the major role 
that these will play in the following developments, it is necessary to enter in some
detail from the outset.
As anticipated in the
Introduction, we shall consider PBC and APBC (precisely cylindrical 
antiperiodic BC) implemented by adding to the interaction an extra term
${\cal B}(\boldsymbol{s})$ with couplings among spins on the boundary~\cite{Gallavotti,Antal,FCZ}.
In the $d=2$ case spins on opposite edges are coupled ferromagnetically, just 
like spins in the bulk, if PBC are applied.
Instead, in the APBC case, spins on one pair of opposite edges
are coupled ferromagnetically, while those on the other pair antiferromagnetically. Hence,
the boundary term reads
\be
{\cal H}_{b}(\boldsymbol{s}) = 
-J \sum_{y=1}^L s_{1,y} s_{L,y} - b J \sum_{x=1}^L s_{x,1}s_{x,L},
\label{bdr.1}
\ee
where we have denoted by 
$b=\pm$ the sign of $J$, which identifies 
PBC $(+)$ or APBC $(-)$.
In the $d=1$ case this term simplifies to
 \be
{\cal H}_b (\boldsymbol{s}) = 
- bJ s_{1}s_{L},
\label{bdr.2}
\ee
where $L$ is the length of the chain. It is important to note that both these BC preserve
the up-down symmetry of the Ising interaction. 

Taking as it is customary $T_I = \infty$ in order to have an uncorrelated initial state,
the system is put in contact with a thermal reservoir at the lower and
finite temperature $T_F$ 
and let to evolve according to a dynamical rule which does not conserve the
order parameter, like Glauber or Metropolis. This simply corresponds to
running a Markov chain at the fixed temperature $T_F$, with the so-called hot start,
that is with a uniformly random initial condition.
The relaxation process is monitored through the equal-time spin-spin correlation function
\be
\mathcal{C}(r, \epsilon,t^{-1},L^{-1}; b)  =   \left [ \langle s_{i}(t)s_{j}(t)\rangle - 
\langle s_{i}(t) \rangle \langle s_{j}(t) \rangle \right ],
\label{corr.0}
\ee
where the angular brackets denote averages taken over the noisy dynamics
and the initial conditions, while the square brackets stand for the average over all
pairs of sites $(i,j)$ keeping fixed the distance $r$ between $i$ and $j$.
In the set of control parameters, $\epsilon=T_F-T_c$ is the temperature difference from
criticality, $t^{-1}$ is the inverse time and $L^{-1}$ is the inverse linear size.

We are interested in taking both
the large-time and the thermodynamic limit
of the above quantity, and then to compare the outcomes,
depending on the order in which these two limits have been taken. Letting
$t^{-1} \to 0$ first, while keeping $L$ fixed, 
the equilibrium correlation function is obtained
\be
\lim_{t^{-1} \to 0}\mathcal{C}(r,\epsilon,t^{-1},L^{-1};b)  =   
\mathcal{C}_{\rm eq}(r,\epsilon,L^{-1};b)  = 
\left [ \langle s_{i}s_{j}\rangle_{\rm eq} - 
\langle s_{i} \rangle_{\rm eq} \langle s_{j} \rangle_{\rm eq} \right ],
\label{D3.7}
\ee
where now the angular brackets stand for the Gibbs ensemble average 
and the square brackets have the same meaning as in
Eq.~(\ref{corr.0}). Then, the subsequent
thermodynamic limit implements the prescription~\cite{Gallavotti} for the construction of
the equilibrium correlation function in the infinite system
\be
\lim_{L^{-1} \to 0} \lim_{t^{-1} \to 0}\mathcal{C}(r,\epsilon,t^{-1},L^{-1};b) =
C_{\rm eq}(r,\epsilon;b).
\label{D3.6}
\ee
The crux of the matter is that, after reversing the order of these 
limits, the end result might not be the same as the one above, 
because the large-time limit of the
time-dependent correlation function for the infinite system
\be
\lim_{t^{-1} \to 0} \lim_{L^{-1} \to 0} \mathcal{C}(r,\epsilon,t^{-1},L^{-1};b) =
C^*(r,\epsilon;b),
\label{D3.5}
\ee
exists but does not necessarily coincide with 
$C_{\rm eq}(r,\epsilon;b)$.
Referring to $C^*(r,\epsilon;b)$ as the time-asymptotic correlation 
function, if it matches $C_{\rm eq}(r,\epsilon;b)$ then
the infinite system equilibrates. If not, it remains permanently out
of equilibrium.
Which is the case depends on $T_F$ and on the choice of BC.
In the
quench to $T_F \geq T_c$ both $C_{\rm eq}(r,\epsilon)$ and 
$C^*(r,\epsilon)$ are independent of the BC choice and do coincide,
signaling equilibration. Instead, in the
quench to below $T_c$, as we shall see, 
$C^*(r,\epsilon)$ does not depend on
$b$, while $C_{\rm eq}(r,\epsilon;b)$ retains this dependence,
implying that equilibration can be achieved at most with one of the two BC, but
certainly not with both. As anticipated 
in the Introduction, the equilibration condition is fulfilled with APBC, but
not with PBC. 

In the next section we shall substantiate the above statements with results for the
$1d$  and $2d$ Ising model.
We shall take the aforementioned limits, after setting up the
general scaling scheme which unifies static and dynamic phenomena
into one single framework encompassing both. 
In order to do this, it is convenient to treat separately the three cases:
$\epsilon > 0$, $\epsilon = 0$ and $\epsilon < 0$.

\section{Statics and Dynamics: $\boldsymbol{\epsilon > 0}$}
\label{III}

\begin{figure}[ht]
\centering
\includegraphics[width=8cm]{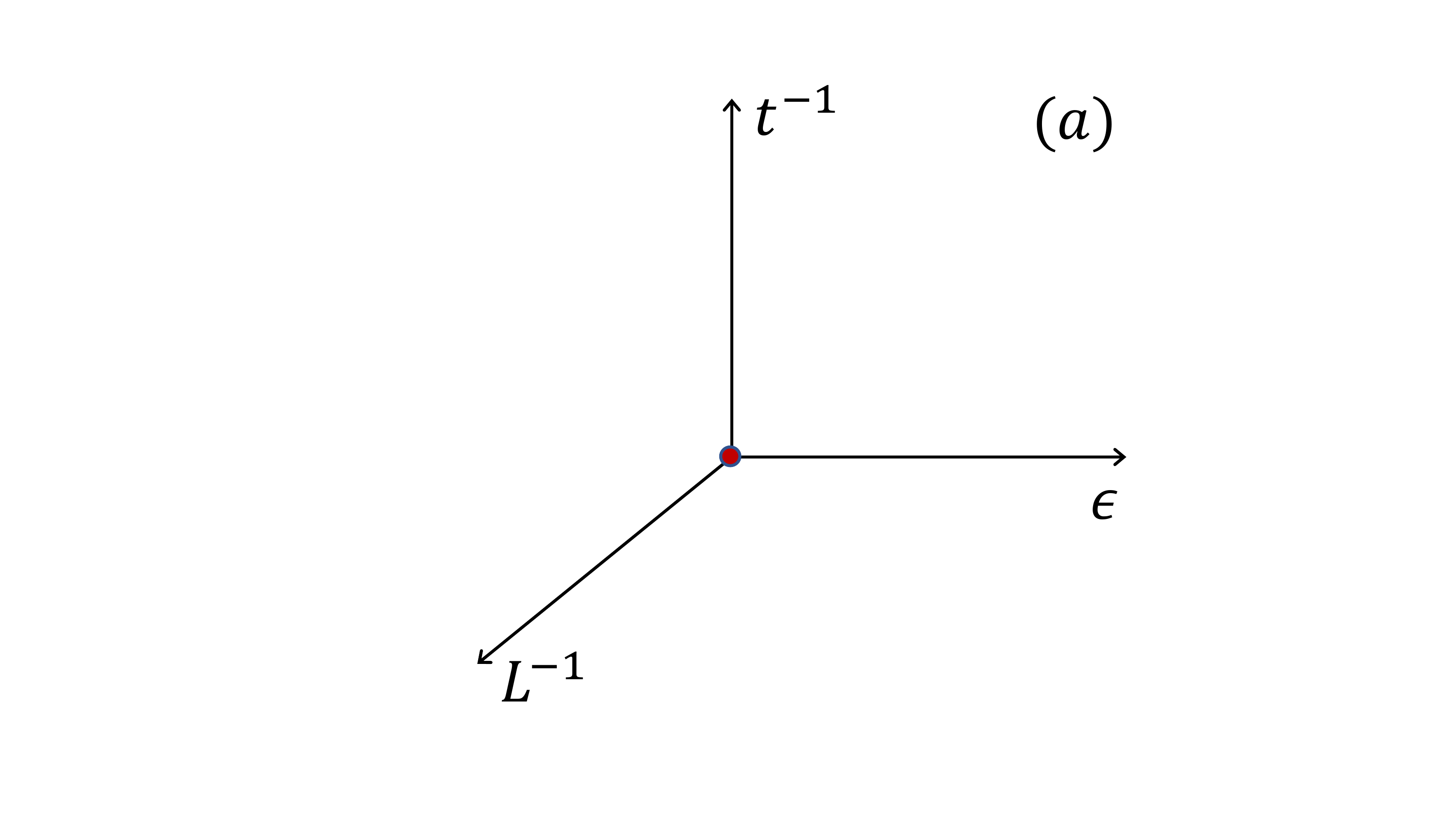} \hspace{-2cm}
\includegraphics[width=8cm]{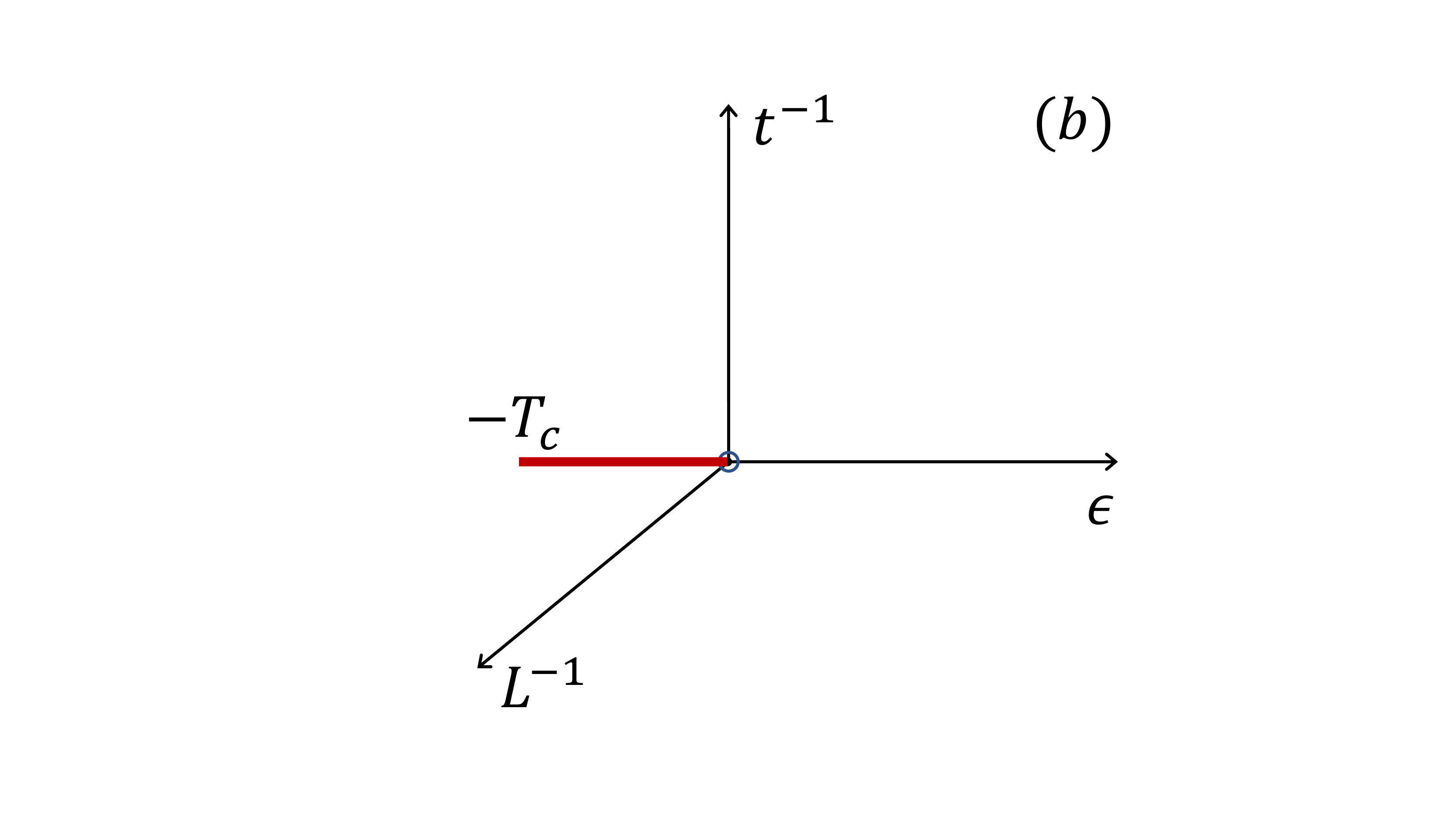}
        \caption{Parameter space of the $1d$ model (a)
        and of the $2d$ model (b).}
\label{fig:pspace}
\end{figure}

\begin{figure}[ht]
\centering
\includegraphics[width=8cm]{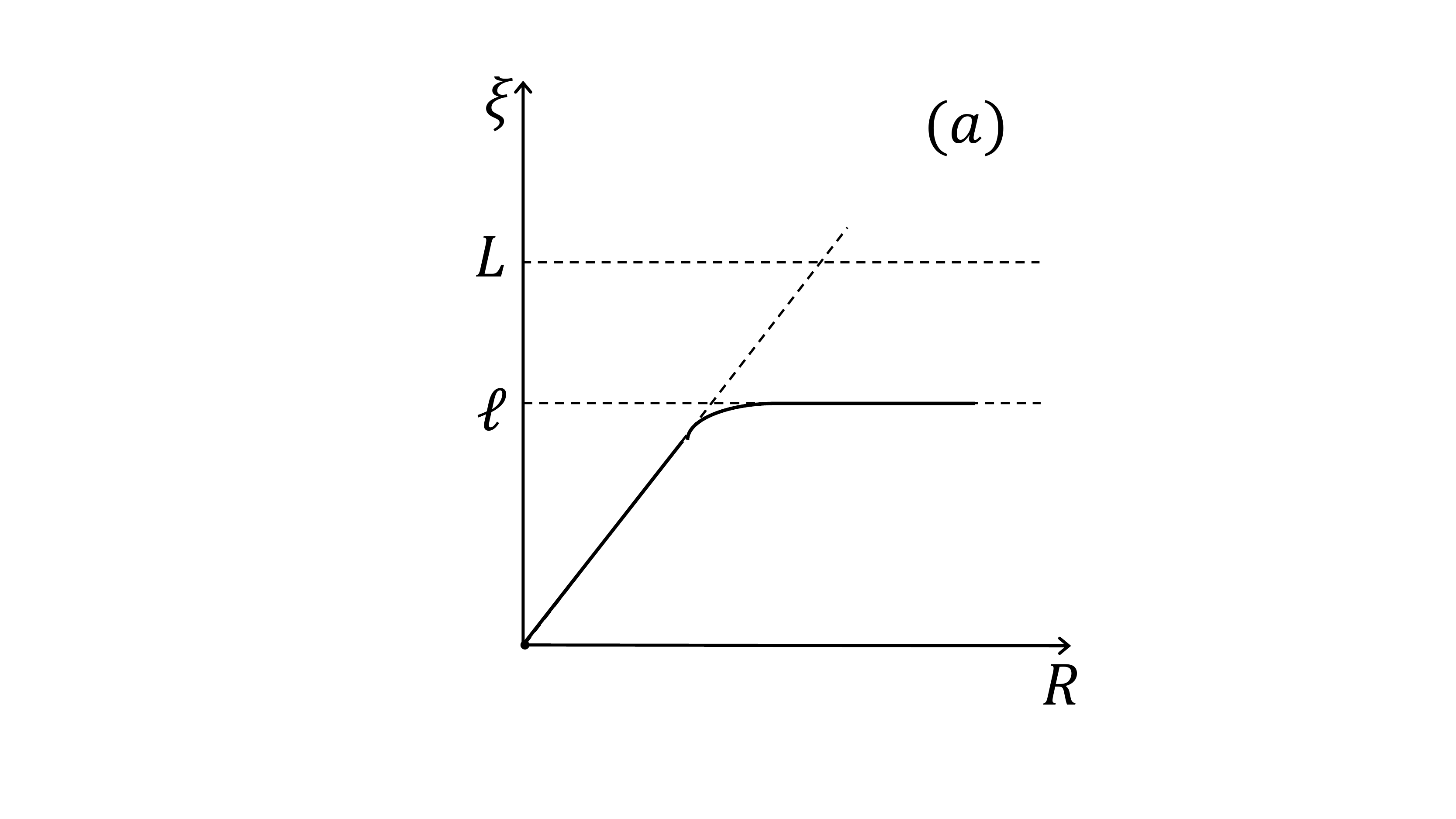} \hspace{-2cm}
\includegraphics[width=8cm]{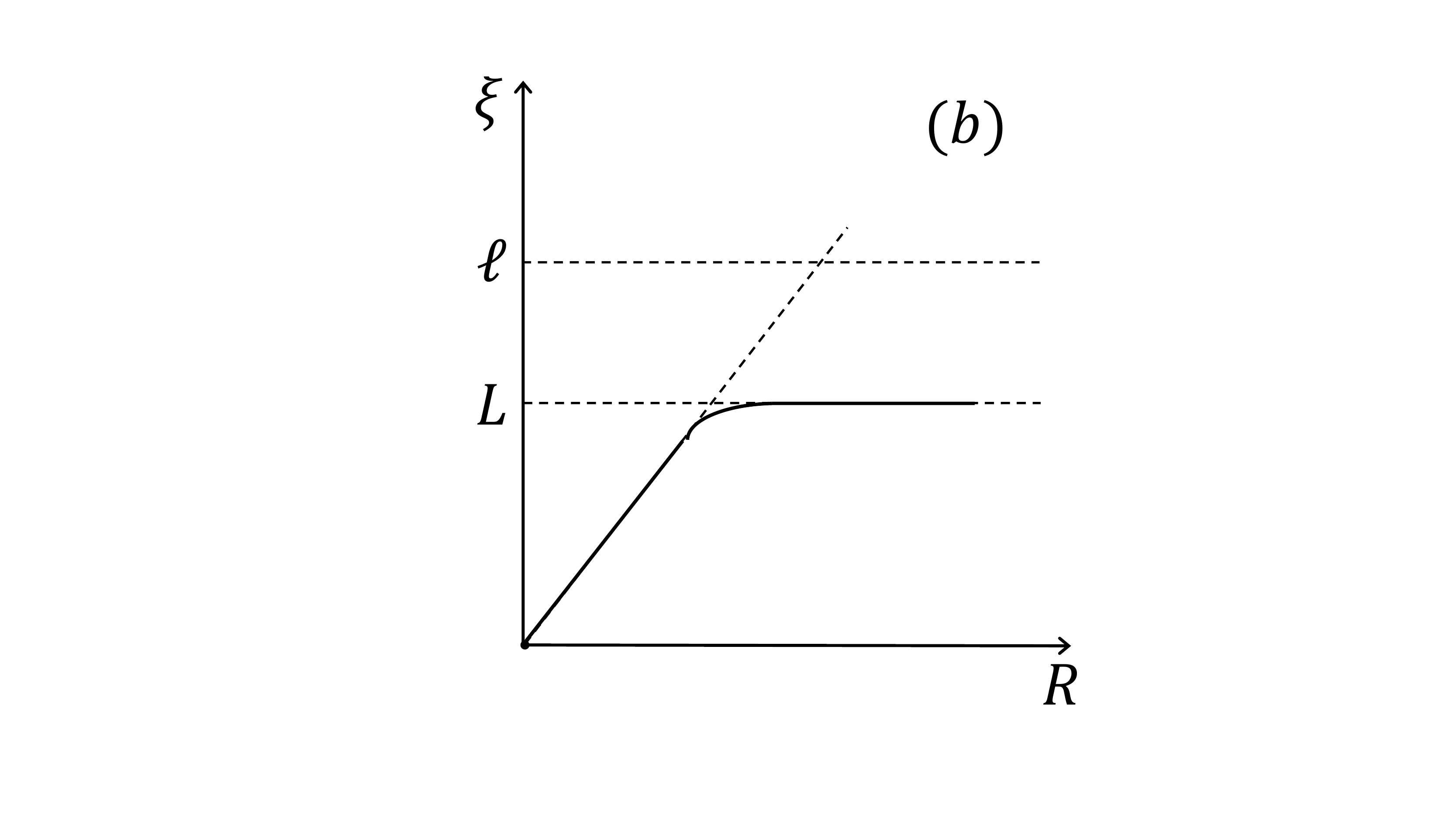}
        \caption{Schematic representation of the
        saturation of $\xi$ vs $R$ for $\ell \ll L$ (a)
        and for $\ell \gg L$ (b).}
\label{fig:R}
\end{figure}

Let us assume that at a generic point in the $\epsilon > 0$ sector
of the three-dimensional space of the parameters
$(\epsilon,t^{-1},L^{-1})$, depicted in Fig.\ref{fig:pspace},
the correlation function obeys scaling in the form~\footnote{This is a finite-size 
extension of the scaling form derived in Ref.~\cite{Janssen}.}
\be
\mathcal{C}(r,\epsilon,t^{-1},L^{-1};b)
= \frac{1}{r^a} \mathcal{F} \left ( \frac{r}{\ell}, \frac{\ell}{R},\frac{R}{L};b \right ),
\label{anml.1}
\ee
where the exponent $a$ is related to the anomalous dimension exponent $\eta$
by $a = d-2+\eta$ and to the fractal dimensionality $D$ of the 
Coniglio-Klein (CK)~\cite{CK,FK} correlated clusters~\cite{CF} by
\be
a = 2(d-D).
\label{anml.2}
\ee
From the exact results~\cite{Stanley,Goldenfeld}
\be
\eta = \left \{ \begin{array}{ll}
         1, \;\; $for$ \;\; d=1,\\
         1/4,\;\; $for$ \;\; d=2, 
        \end{array}
        \right .
        \label{z.0}
        \ee
follows
\be
a = \left \{ \begin{array}{ll}
         0, \;\; $for$ \;\; d=1,\\
         1/4,\;\; $for$ \;\; d=2, 
        \end{array}
        \right .
        \label{z.1}
        \ee
and
 \be
D = \left \{ \begin{array}{ll}
         1, \;\; $for$ \;\; d=1,\\
         15/8,\;\; $for$ \;\; d=2, 
        \end{array}
        \right .
        \label{z.2}
        \ee  
which shows that the CK clusters are compact in $1d$ and fractals in $2d$.
Up to a proportionality constant, the scaling variable $\ell$ 
is the equilibrium correlation length of the infinite system, given by~\cite{Stanley}
\be
\ell(\epsilon)
= \left \{ \begin{array}{ll}
         -[\ln \tanh (J/\epsilon)]^{-1}, \;\; $for$ \;\; d=1,\\
         \epsilon^{-\nu},\;\; $with$ \;\; \nu=1 \;\; $for$ \;\; d=2. 
        \end{array}
        \right .
        \label{anml.3}
        \ee
The other characteristic length $R(t)$ obeys the power law~\cite{Janssen}        
\be
R(t) = t^{1/z},
\label{anml.3bis}
\ee
with the dynamical exponent~\cite{1d,Nightingale}
\be
z
= \left \{ \begin{array}{ll}
         2, \;\; $for$ \;\; d=1,\\
         2.16,\;\; $for$ \;\; d=2.  
        \end{array}
        \right .
        \label{z.1}
        \ee
The connection between $R(t)$ and the time dependent correlation length will be
clarified shortly and is summarized in Fig.\ref{fig:R}.  
Both lengths diverge as the critical point, which is at the origin of the 
reference frame in Fig.\ref{fig:pspace}, is 
approached along the $\epsilon$ axis and the $t^{-1}$ axis, respectively.

The scaling ansatz~(\ref{anml.1}) is dense of information and allows to predict what
should be expected in different regions of the parameter space. The foremost relevant
features are the power-law decay $1/r^{a}$ of correlations at 
short distance and the large-distance cutoff enforced by the scaling
function. The separation between short and large distances is fixed by the
correlation length 
\be
\xi(\epsilon,t^{-1},L^{-1};b) = 
\left [ \frac{\int_0^L dr \, r^2 \, \mathcal{C}(r,\epsilon,t^{-1},L^{-1};b) }
{\int_0^L dr \,  \mathcal{C}(r,\epsilon,t^{-1},L^{-1};b) } \right ]^{1/2},
\label{corrl.0}
\ee
which scales as
\be
\xi(\epsilon,t^{-1},L^{-1};b) = R f \left ( \frac{R}{\ell}, \frac{\ell}{L};b \right ).
\label{crrl.1}
\ee
The behaviour of $\xi$, as parameters are changed,
can be unraveled by the following argument. Suppose that $\ell$ and $L$ are fixed
in a region where $\ell \ll L$ and let us survey what happens
as the quench unfolds and $R$ grows.
Approximating the above equation by
\be
\xi(\epsilon,t^{-1}) \simeq R f \left ( \frac{R}{\ell},0 \right ),
\label{crrl.101}
\ee
in the early stage of the quench, when $R \ll \ell$, it can be further reduced to 
\be
\xi \sim R,
\label{xiR.1}
\ee
because the system behaves as if it was approaching the critical
point along the $t^{-1}$ axis. As $R$ is let to grow further, equilibrium is 
eventually reached 
when $R \sim \ell$ and the correlation length saturates to the limiting
value
\be
\xi \sim \ell,
\label{elle.1}
\ee 
as illustrated in the left panel of Fig.\ref{fig:R}, with the equilibration time
given by $t_{\rm eq} = \ell^z$.
The BC are immaterial throughout, because $\xi$ remains always much smaller than $L$
so that the system as a whole behaves as a collection of independent
finite systems, on which the far away BC have no effect. 
By the same reasoning, in the region where $\ell \gg L$ we still have $\xi \sim R$
in the early stage, when $R \ll L$, with independence from BC. But then BC come into
play when the system equilibrates and
$\xi$ saturates to the limiting value
$L$, as illustrated in the right panel of Fig.\ref{fig:R},
since correlations extend up to distances where the BC are effective.
In this connection see Ref.~\cite{Das}.
Summarizing, $\xi$ is given by the shortest of the three lengths
$(\ell, R,L)$, that is
\be
\xi(\epsilon,t^{-1},L^{-1}) \sim \min(\ell,R,L),
\label{crrl.4}
\ee 
in the regions of the parameter space where one of the three is considerably 
shorter than the other two,
with crossovers connecting these regions. 
It is clear from Fig.\ref{fig:R} that $\xi$ and $R$ coincide at 
all times if both $\ell$ and $L$ are infinite.
Next to the correlation length, it is useful to keep track also of the
susceptibility
\be
\chi(\epsilon,t^{-1},L^{-1};b) = 
\int d^d r \,  \mathcal{C}(r,\epsilon,t^{-1},L^{-1};b),
\label{susc.0}
\ee
which is related to the correlation length by
\be
\chi \sim \xi^{2D-d}.
\label{susc.1}
\ee
This is an important relation, because it is independent of the direction of approach
to the critical point and depends only on the geometrical nature of the correlated clusters 
through $D$.

According to the above reasoning, when the limits $t^{-1} \to 0$ and $L^{-1} \to 0$ 
are taken in the $\epsilon > 0$ sector, we necessarily have $\xi \sim \ell$,
independently of the order in which these limits are taken, because $\ell$
is finite. Moreover, the finite correlation
length guarantees that the system equilibrates
with independence from BC 
\be
\lim_{L^{-1} \to 0} \lim_{t^{-1} \to 0} 
\mathcal{F} \left ( \frac{r}{\ell}, \frac{\ell}{R},\frac{R}{L};b \right ) =
\lim_{t^{-1} \to 0} \lim_{L^{-1} \to 0} 
\mathcal{F} \left ( \frac{r}{\ell}, \frac{\ell}{R},\frac{R}{L};b \right ) =
F_{\rm eq} \left (\frac{r}{\ell} \right ).
\label{D3.6bis}
\ee

\subsection{Example: $1d$ system}

As an example, let us check the above statements against exact results
in the particular case of the 
$t^{-1} \to 0$ limit of the $1d$ model with finite $L$. 
The equilibrium correlation function is given by 
\be
\mathcal{C}_{\rm eq}(r,\epsilon,L^{-1};b)
= \frac{1}{r^a} \mathcal{F}_{\rm eq} \left ( \frac{r}{L}, \frac{L}{\ell};b \right ),
\label{anml.4}
\ee
where $a=0$ according to Eq.~(\ref{z.1}), while
the two explicit forms of the scaling function (see I and Ref.~\cite{Antal}) read
\be
\mathcal{F}^{(p)}_{\rm eq}(z,\zeta) = \frac{\cosh[\zeta(1-z)]}{\cosh(\zeta)},
\label{anml.6}
\ee
\be
\mathcal{F}^{(a)}_{\rm eq}(z,\zeta) = \frac{\sinh[\zeta(1-z)]}{\sinh(\zeta)},
\label{anml.7}
\ee
where we have set
\be
z=r/L, \quad \zeta = L/\ell.
\label{anml.8}
\ee
We have considered a chain of length $2L$ in order to simplify notation.
The superscripts $(p)$ and $(a)$ have been used for PBC and for APBC, respectively.
The equilibrium correlation length, defined through the second moment as 
in Eq.~(\ref{corrl.0}),
scales as  
\be
\xi_{\rm eq}(\epsilon,L^{-1};b) = \ell f_{\rm eq}(\zeta;b),
\label{crrl.1bis}
\ee
with the scaling functions 
\be
 f_{\rm eq}^{(p)} (\zeta) =  \left [2-\frac{2\zeta}{\sinh (\zeta)} \right ]^{1/2},
 \quad f_{\rm eq}^{(a)} (\zeta) =  \left [2-\frac{\zeta^2}{\cosh(\zeta) - 1} \right ]^{1/2},
\label{crrl.5}
\ee
from which follows 
\be
\xi_{\rm eq}^{(p)}(\epsilon,L^{-1}) =  \left \{ \begin{array}{ll}
         \sqrt{2} \, \ell, \;\; $for$, \;\; \ell \ll L,\\
         \frac{1}{\sqrt{3}} \, L,\;\; $for$, \;\; L \ll \ell, 
        \end{array}
        \right .
        \quad 
\xi_{\rm eq}^{(a)}(\epsilon,L^{-1}) =  \left \{ \begin{array}{ll}
         \sqrt{2} \, \ell, \;\; $for$, \;\; \ell \ll L,\\
         \frac{1}{\sqrt{6}} \, L,\;\; $for$, \;\; L \ll \ell, 
        \end{array}
        \right .       
        \label{crrl.10}
        \ee
showing that in the regimes $\ell \ll \L$ and $\ell \gg L$, indeed
one has $\xi \sim \min(\ell,L)$.
Completing, next, the sequence of limits by letting
$L^{-1} \to 0$, it is straightforward to check that the dependence on BC disappears,
yielding
\be
\lim_{L^{-1} \to 0} \mathcal{F}^{(p)}_{\rm eq}(z,\zeta)
= \lim_{L^{-1} \to 0} \mathcal{F}^{(a)}_{\rm eq}(z,\zeta) = e^{-r/\ell}.
\label{anml.9}
\ee
Using the definition, it is immediate to verify that also $\chi \sim \ell$ and, 
therefore, that moving toward the critical point along the $\epsilon$ axis one has
\be
\chi \sim \xi,
\label{susc.2}
\ee
in agreement with Eq.~(\ref{susc.1}), because $D=1$ when $d=1$.

\section{Statics and Dynamics: $\boldsymbol{\epsilon = 0, d=2}$}
\label{IV}

When $\epsilon = 0$, the $1d$ and $2d$ cases are quite different and need 
to be treated separately. In the latter one, which we shall now consider,
$T_c > 0$ and ergodicity does not break. In the former, instead, $T_c = 0$
and ergodicity may break, depending on BC. This makes it more akin to 
the $2d$ quench to below $T_c$. So, it will be dealt with in the next section.

\begin{figure}[ht]
\centering
\includegraphics[width=10cm]{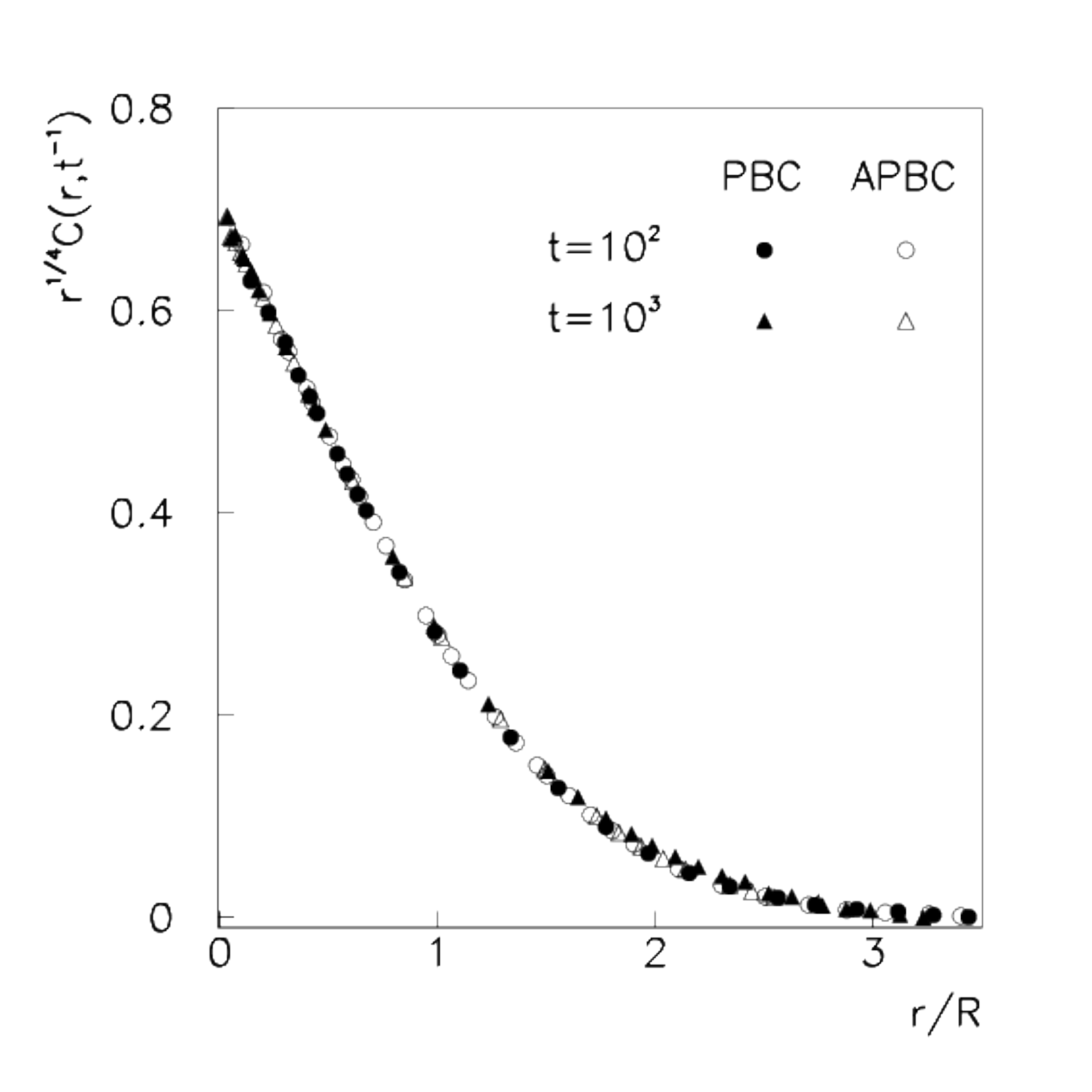}
        \caption{Scaling function of the time-dependent correlation
        function in the quench to $T_c$ of the $2d$ model, 
        demonstrating independence from BC, in the system with $L=256$.
        PBC (black symbols) and APBC (empty symbols).}
\label{fig:collapseTc}
\end{figure}

The specificity of the quench to $\epsilon = 0$ is that $\ell$ diverges and, consequently,
that $\xi$ can be limited only by $R$ or $L$. Thus, when the $t^{-1} \to 0$ limit is taken first
and $L$ is kept fixed, 
$\xi$ crosses over from $R$ to $L$ in the finite time $t_{\rm eq} \sim L^z$,
as in the right panel of Fig.\ref{fig:R}, and the system equilibrates to
\be
\mathcal{C}_{\rm eq}(r,L^{-1};b) = \frac{1}{r^{1/4}} 
\mathcal{F}_{\rm eq} \left ( \frac{r}{L};b \right ),
\label{sat.1}
\ee
which depends on BC because correlations extend up to the boundary.
Letting next $L^{-1} \to 0$, the BC dependence disappears from the 
critical correlation function of the infinite system 
\be
C_{\rm eq}(r) \sim \frac{1}{r^{1/4}}.
\label{sat.2}
\ee

Instead, if the $L^{-1} \to 0$ limit is taken first, $R$ is the only length
left in the problem. This implies $\xi \sim R$ at all times,  so that there is no
finite equilibration time. However, the time-dependent
correlation function 
\be
C(r,t^{-1}) = \frac{1}{r^{1/4}}  F_c (r/R),
\label{sat.3}
\ee
which is BC independent (see Fig.\ref{fig:collapseTc}), 
gets arbitrarily close to the equilibrium counterpart~(\ref{sat.2}) as time 
grows, because the limit
\be
\lim_{t^{-1} \to 0} C(r,t^{-1}) = C^*(r) \sim \frac{1}{r^{1/4}},
\label{sat.4}
\ee
coincides with it. 
In summary, in the quench to $\epsilon = 0$ of the $2d$ system, like in
the $\epsilon > 0$ case previously considered, the equilibrium correlation
function of the infinite system $C_{\rm eq}(r)$ does not depend on
BC and coincides with the time-asymptotic one $C^*(r)$,
warranting the conclusion that the system can get arbitrarily close to
equilibrium by waiting long enough.

Comparing Eqs.~(\ref{sat.1}) and~(\ref{sat.3}), it is evident that the scaling
structure is the same, the only difference being in the specific forms of the
scaling functions, which is inessential for the present considerations. This shows
that the time direction along the $t^{-1}$ axis, as far as scaling is concerned, 
is just another direction
of approach to the critical point, on the same footing with the other two.
In addition, from the formal similarity of the two scaling expressions follows
straightforwardly that the susceptibility satisfies Eq.~(\ref{susc.1}) in the form
\be
\chi \sim \xi^{7/4},
\label{suscett.1}
\ee
irrespective of the direction of approach, with $\xi \sim L$ along the $L^{-1}$
axis and $\xi \sim R$ along the $t^{-1}$ axis.

\section{Statics and Dynamics: $\boldsymbol{\epsilon = 0, d=1}$}
\label{V}

\begin{figure}[ht]
\centering
\includegraphics[width=5cm]{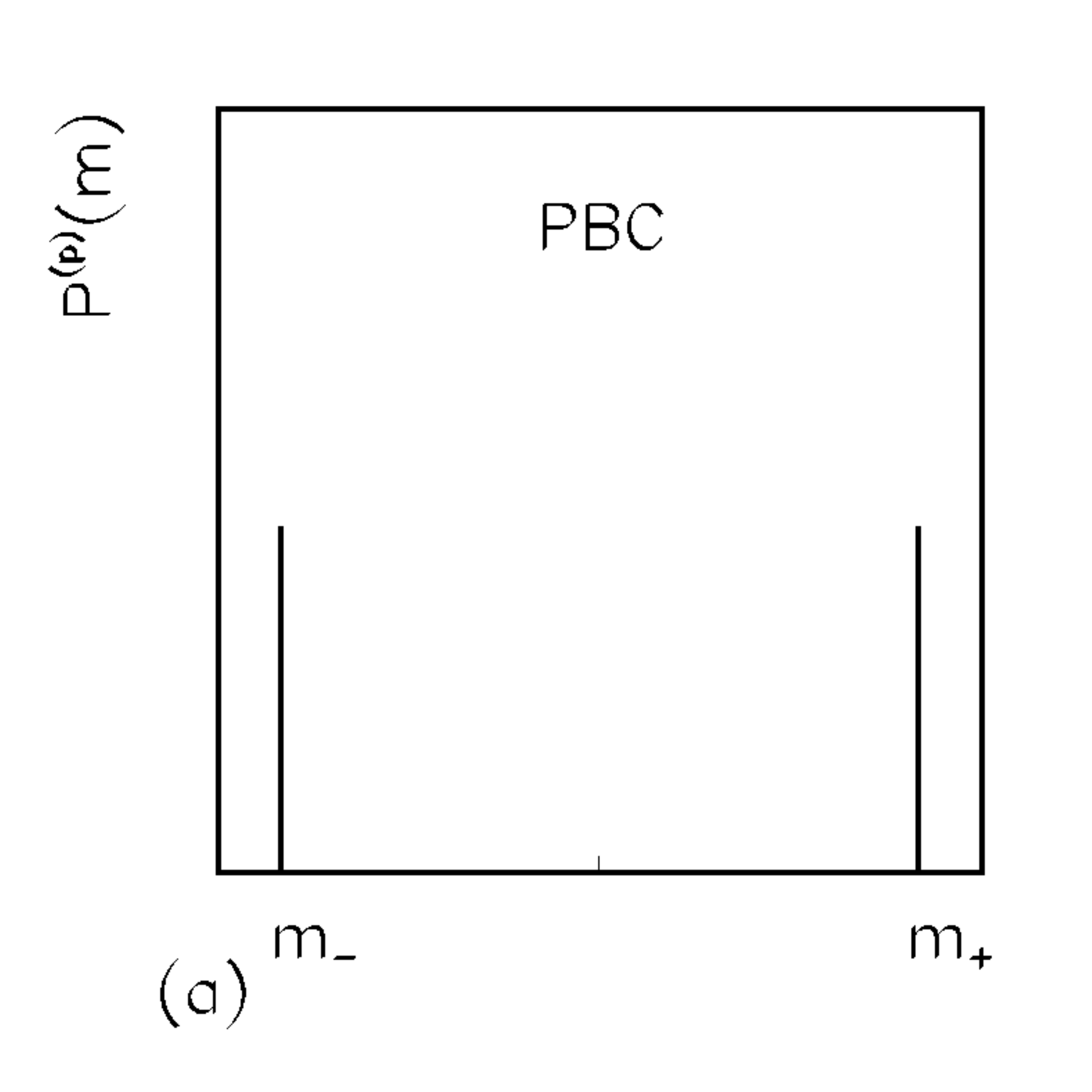}
\includegraphics[width=5cm]{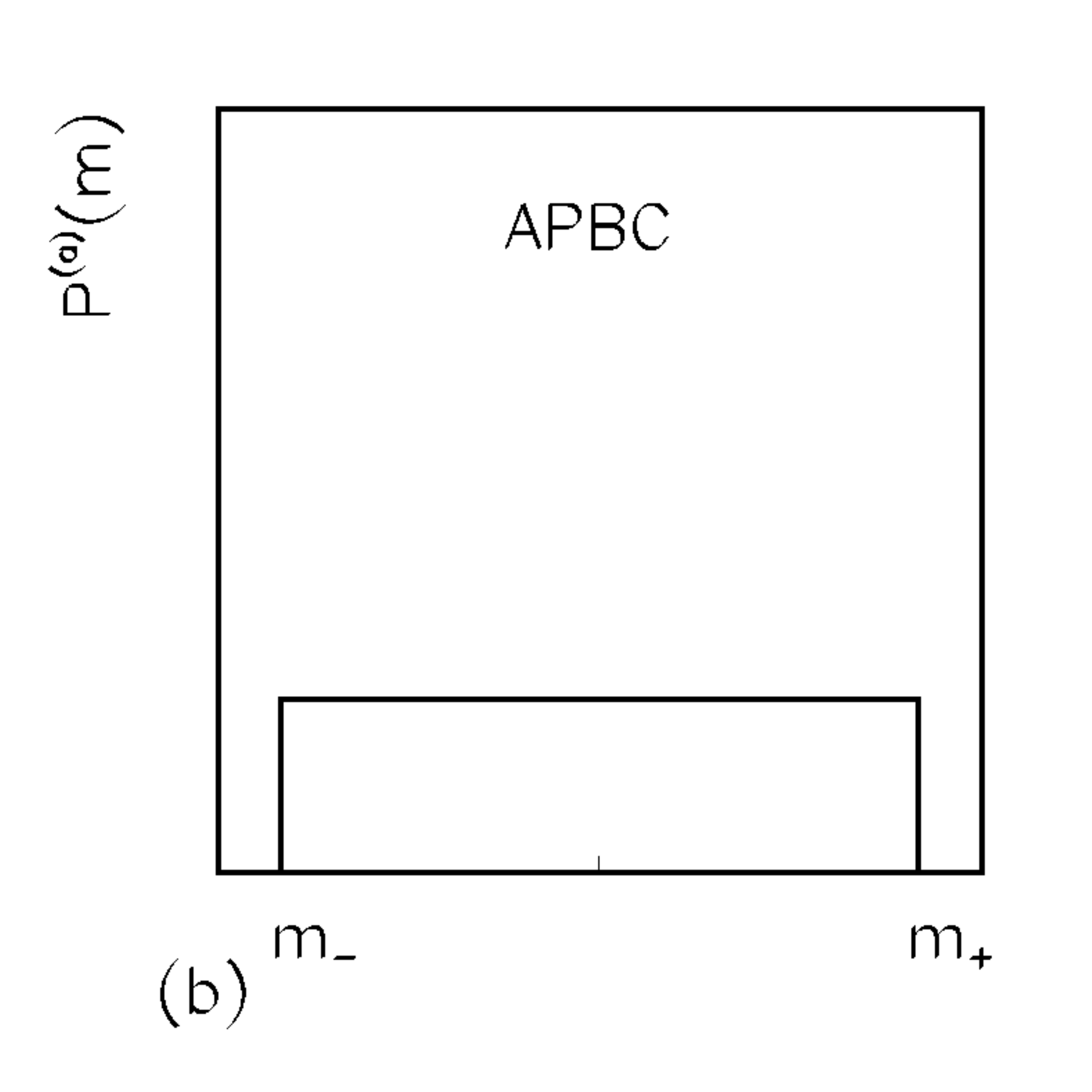}
\caption{Magnetization density distributions at $\epsilon=0$ in the $1d$ Ising model,
	with $m_\pm = \pm 1$. The spikes in the panel (a) stand for $\delta$ functions.} 
\label{fig:1}
\end{figure}

As mentioned above and explained at length in I,
in the $1d$ system at $\epsilon = 0$ we are confronted with a radically different situation,
because ergodicity, which holds for both BC above $T_c$, is now broken with PBC, 
but not with APBC. In order to ease the comparison, and to highlight 
the contrast, with the less familiar case of a transition without ergodicity breaking,
let us first briefly summarize the well-established concept of ergodicity breaking~\cite{Palmer}. In the 
PBC case there are two degenerate ground states: the two ordered configurations
with all spins either up $\boldsymbol{s}_+ = [s_i=+1]$ or
down $\boldsymbol{s}_- = [s_i=-1]$. These, by themselves, form
two absolutely-confining ergodic components, which are
dynamically disconnected because the activated moves  needed to go from one to the
other are forbidden at zero temperature. Consequently, time averages 
coincide with ensemble averages taken with either one of the two
broken-symmetry ferromagnetic pure states
$P_-(\boldsymbol{s})=\delta_{\boldsymbol{s},\boldsymbol{s}_-}, P_+(\boldsymbol{s})=\delta_{\boldsymbol{s},\boldsymbol{s}_+}$
and {\it do not} coincide with the symmetric 
ensemble averages taken in the Gibbs state, which is the
even mixture of the pure states
\be
P^{(p)}(\boldsymbol{s}) = \frac{1}{2}[P_-(\boldsymbol{s}) + P_+(\boldsymbol{s})].
\label{mixt.0}
\ee
In such a situation, only time averages are physically meaningful.
Conversely, in the APBC case all the $4L$ degenerate ground-state 
configurations with one defect (or domain wall) belong to the same
ergodic component, because the defect can freely sweep 
the whole system at no energy cost. Then, in this case 
time and ensemble averages coincide.
The qualitative difference between the two zero-temperature states is 
well illustrated (see Fig.\ref{fig:1}) by the
probability distribution $P_b(m)$ of the magnetization 
density $m=\frac{1}{2L}\sum_i s_i$, which is demonstrated~\cite{Antal} to be
double peaked in the PBC case
\be
P^{(p)}(m) = \frac{1}{2} [\delta(m+1) + \delta(m-1)],
\label{mxt.1}
\ee
and uniform over the $[-1,1]$ interval in the APBC case
\be
P^{(a)}(m) \to  \left \{ \begin{array}{ll}
         1/2, \;\; $for$, \;\; m \in [-1,1],\\
         0,\;\; $for$, \;\; m \notin [-1,1].  
        \end{array}
        \right .
        \label{mxt.2}
        \ee
        
So, if we now take the $t^{-1} \to 0$ limit while keeping $L^{-1}$ fixed, 
we find  BC-related differences in the results.
With PBC, as explained above, the meaningful averages are those 
in the broken symmetry pure states, yielding
\be
\mathcal{C}^{(p)}_{\rm eq}(r,L^{-1}) = [\langle s_is_j \rangle_\pm - \langle s_i \rangle_\pm \langle s_j \rangle_\pm] = 0,
\label{1dI.0}
\ee        
where the angular brackets stand for
the average with respect  to $P_-(\boldsymbol{s})$ or $P_+(\boldsymbol{s})$. 
The vanishing of correlations
for any $r$ holds independently of $L$ and clearly implies
that also the correlation length vanishes. 
Notice that
if the correlation function had been extracted by taking the $\epsilon \to 0$ limit 
of the Gibbs average taken with the distribution~(\ref{mixt.0}), as in 
 Eq.~(\ref{anml.6}), the result would have been
\be
\mathcal{C}^{(p)}_{\rm eq}(r,L^{-1}) = 1,
\label{1dI.16}
\ee        
which is independent of $r$ and does not decay. However, this would have been 
just an artefact of the mixture.

Conversely, in the APBC case the ensemble Gibbs average, as calculated
in Eq.~(\ref{anml.7}), gives the correct time-average result
because ergodicity is not broken. Hence, in the $\epsilon \to 0$ limit,
from Eq.~(\ref{anml.7}) one has
\be
\mathcal{C}^{(a)}_{\rm eq}(r,L^{-1}) = (1-r/L).
\label{1dI.17}
\ee
The dependence on $r/L$ in the above expression reveals 
that correlations extend over a distance of order L in agreement with the
general argument expounded in section~\ref{III}.
Hence, by letting $L^{-1} \to 0$  the correlation length diverges, leading
to the conclusion that
the state at the origin of the parameter space
is a {\it critical point} for the APBC system, where the correlation function
displays the constant behaviour
\be
C^{(a)}_{\rm eq}(r) = 1.
\label{corrl.01}
\ee
Contrary to Eq.~(\ref{1dI.16}), now the lack of decay is a real physical effect, which corresponds to the critical power law decay $1/r^{a}$ with a vanishing
exponent $a$, due to the compactness of the CK 
correlated clusters.

\begin{figure}[ht]
\centering
\includegraphics[width=10cm]{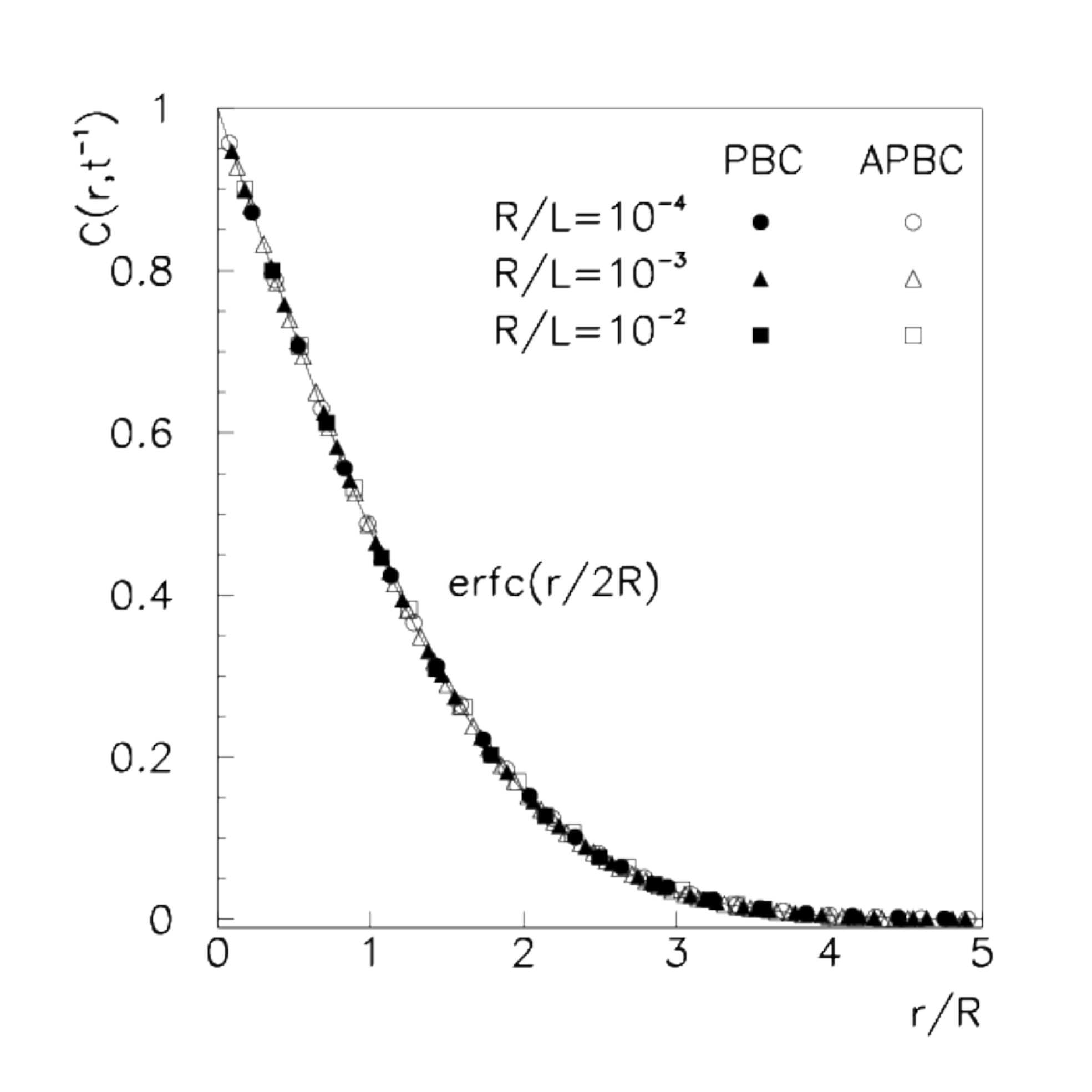}
\caption{Collapse on the master curve of Eq.~(\ref{quench.1}) of the data for
$C(r,t^{-1},L^{-1})$ in the time regime $R \ll L$. PBC (black symbols)
and APBC (empty symbols). The data for $R/L=10^{-4},10^{-3}$ have been obtained with 
$L=10^5$, those with $R/L=10^{-2}$ with $L=10^4$.}
\label{fig:scaledC}
\end{figure}

When the sequence of limits is reversed, after taking the thermodynamic
limit we are again in the situation in which $R$ is
the only length in the problem. Therefore $\xi \sim R$, as in the previous section,
and we get the BC independent result
\be
C(r,t^{-1}) = \frac{1}{r^a} F (r/R ),
\label{q.1}
\ee
with $a=0$. The function $F(x)$
is known from exact analytical computation with 
PBC~\cite{1d,Bray} and is given by
\be
F(x) = \erfc(x), \quad \text{with} \quad x=r/2R.
\label{quench.1}
\ee
That the same scaling function applies also to the case of APBC is demonstrated
by the numerical data displayed in Fig.~\ref{fig:scaledC}, which have been
obtained by simulating the quench dynamics with the
Metropolis algorithm on a system with $L=10^5$, after imposing PBC and APBC.
The plot shows that the above result indeed holds irrespective of the
BC choice, because the PBC and APBC data superimpose to the theoretical
curve of Eq.~(\ref{quench.1}) with great accuracy, as long as
$R(t) \ll L$. 
The existence
of an endlessly growing correlation length $R(t)$ means that the relaxation dynamics
along the $t^{-1}$axis drives both systems, with PBC and with APBC, 
toward the same asymptotic critical state at the origin 
with the unique time-asymptotic correlation function given by
\be
\lim_{t^{-1} \to 0} C(r,t^{-1}) = C^*(r)= 1,
\label{elle.2}
\ee
which coincides with the APBC equilibrium result in Eq.(\ref{corrl.01}). 
So, if we compare the asymptotic result of Eq.~(\ref{elle.2})
with the APBC static one of Eq.~(\ref{corrl.01}), 
and with the PBC equilibrium result of Eq.~(\ref{1dI.0}), 
we see, as stated in the Introduction, 
that the APBC system tends toward equilibrium, although with an infinite
relaxation time, while the PBC system remains permanently out of equilibrium.
It is evident that the origin of the diversity of behaviours is in the presence or
absence of ergodicity breaking. In fact, we shall see in the next section that
the same behaviour occurs in the quench of the $2d$ system to below the
critical point.

In order to complete the picture of critical behaviour, let us 
check on the validity of Eq.~(\ref{susc.1}).  From 
Eqs.~(\ref{anml.9},\ref{1dI.17},\ref{quench.1}) follows that along the three
directions one has $\xi \sim \ell, \xi \sim R, \xi \sim L$, as well as 
$\chi \sim \ell, \chi \sim R, \chi \sim L$, yielding
\be
\chi \sim \xi,
\label{susc.10}
\ee
independently of the direction of approach to the critical point, as it should be
since $D=1$.

\section{Statics and Dynamics: $\boldsymbol{\epsilon < 0, 2d}$}
\label{VI}

As in the previous
case, the nature of the equilibrium state
of the $2d$ model below $T_c$ depends strongly on BC, even in the thermodynamic limit. 
In I we have shown
that the segment with $\epsilon < 0$ in the parameter space (see right panel of
Fig.\ref{fig:pspace}) 
is the coexistence line of states spontaneously-magnetized in opposite
directions, when PBC are imposed,
while it is a line of critical points with APBC. Since this is a crucial point, let
us overview the equilibrium picture before turning to the discussion of the
quench dynamics.

\begin{figure}[ht]
\centering
	\includegraphics[width=10cm]{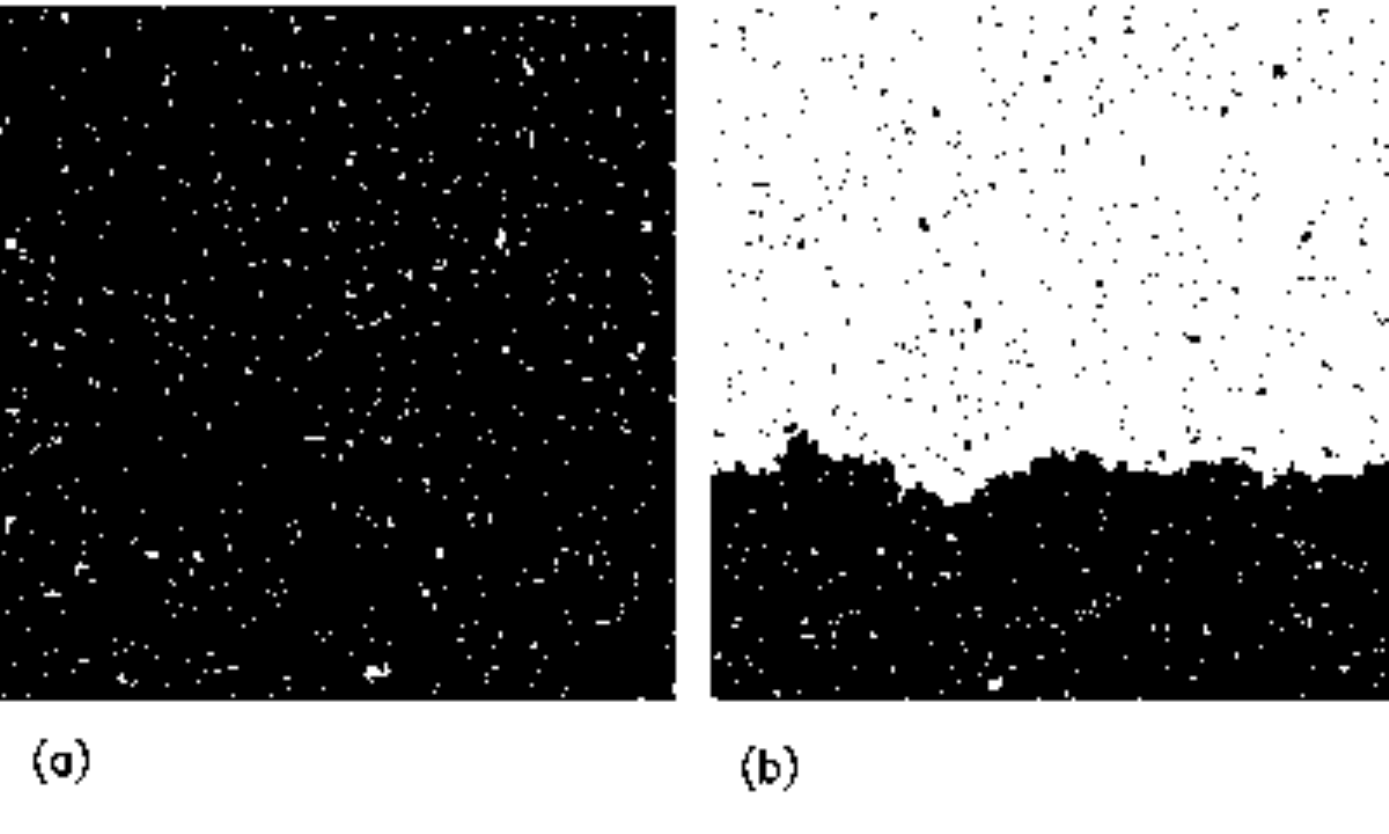}
        \caption{Typical equilibrium configurations below $T_c$. 
        In the PBC case (a) one black domain of up spins fills
        the entire systems. Thermal fluctuations produce the small white domains of
        down spins. In the APBC case (b) there are two large domains separated
        by one interface cutting across the system. Within each domain there are
        the small patches of reversed spins due to thermal fluctuations.}
\label{fig:conf}
\end{figure}

\subsection{{\bf Equilibrium with PBC}}
\label{PBCsotto}

When PBC are imposed, two confining components of spins aligned
either prevalently up or prevalently down, are formed in phase space. 
A configuration typical of the up component 
is shown in the left panel of Fig.\ref{fig:conf}. 
In the thermodynamic limit these components become absolutely
confining, ergodicity breaks down and, therefore, we are confronted with the same
situation discussed in the $1d$ case at $T=0$. Namely,
the Gibbs state becomes
the even mixture of the two broken-symmetry pure states like in Eq.~(\ref{mixt.0}),
that is
\be
P_{\rm eq}^{(p)}(\boldsymbol{s}) = 
\sum_\alpha p(\alpha) P_\alpha (\boldsymbol{s}).
\label{mix.1}
\ee
Here, $\alpha = \pm$ is the component label, the mixing
probability is uniform $p(\alpha)=1/2$ and $P_\alpha (\boldsymbol{s})$ is the
ferromagnetic pure state. The nonvanishing spontaneous 
magnetization density 
$m_\alpha$ is given by
\be
m_-=-m_+,  \quad |m_\alpha| = |\epsilon|^\beta, \quad \beta=1/8.
\label{mix.2}
\ee
Using the above definitions and rewriting the Gibbs average in terms of the component
averages, i.e.  $\langle \cdot \rangle_{\rm eq} = \sum_\alpha p(\alpha)
\langle \cdot \rangle_\alpha$, the equilibrium correlation 
function can be rearranged in the form
\be
C_{\rm eq}^{(p)}(r,\epsilon) = \overline{ \langle (s_i -m_\alpha)
(s_{i+r} -m_\alpha) \rangle_\alpha} \; + \;
\overline{[\langle s_i \rangle_\alpha - \overline{m_\alpha}]
[\langle s_{i+r} \rangle_\alpha - \overline{m_\alpha}]},
\label{mix.4}
\ee
where the overline denotes averaging with respect to $p(\alpha)$.
The first contribution is the average over components of the {\it intra}-component 
correlation function
$\langle \psi_i \psi_{i+r} \rangle_\alpha$, where the variables $\psi_i = s_i - m_\alpha$
represent the thermal fluctuations in the pure state 
$P_\alpha (\boldsymbol{s})$. As it is intuitively clear, deviations from the
average by symmetry do not depend on $\alpha$, so we shall use the notation 
$G_{\rm eq}(r,\epsilon)$ for $\langle \psi_i \psi_{i+r} \rangle_\alpha$. 
At low $T_F$ this quantity is short ranged,
since in the broken-symmetry state the correlation length $\xi_\psi$ 
of the $\psi$ variables vanishes as $T_F \to 0$.
The second term, instead, represents the {\it inter}-components
contribution, which reduces to $m^2_\alpha$, since
$\overline{m_\alpha}=0$ and $m^2_\alpha$ is independent of $\alpha$. Thus,
in the end, from the Gibbs average we have
\be
C_{\rm eq}^{(p)}(r,\epsilon) = G_{\rm eq}(r,\epsilon) +m^2_\alpha.
\label{mix.5}
\ee
It is important, for what follows, to keep in mind that the constant term 
$m^2_\alpha$, which is the variance of the variable $m_\alpha$ distributed according
to $p(\alpha)$, arises exclusively from the mixing as 
the constant term in Eq.~(\ref{1dI.16}).
Therefore, in the PBC case the only dynamical variables 
are the $\psi_i$, which means that the dynamical rule updates $\psi_i$, but not $m_\alpha$.
The magnetization distribution exhibits
the double peak structure~\cite{Binder,Bruce}
which, in the thermodynamic limit, becomes the sum of the two $\delta$ functions
\be
P^{(p)}(m|\epsilon) = \frac{1}{2} [ \delta(m-m_-) + \delta(m-m_+)].
\label{mix.60}
\ee
Hence, as explained in the $1d$ case, the meaningful averages are those taken
with the broken symmetry ensembles $P_\alpha (\boldsymbol{s})$, which coincide with
time averages and give 
\be
C_{\alpha,\rm eq}^{(p)}(r,\epsilon) = G_{\rm eq}(r,\epsilon).
\label{mix.5bis}
\ee

\subsection{{\bf Equilibrium with APBC}}
\label{APBCeq}

When APBC are imposed, like in the $1d$ case ergodicity does 
not break. As explained in I, there is only one ergodic component,
whose typical configurations at sufficiently low $T_F$ are composed of two large
ordered domains, separated by one interface cutting across the system and
sweeping through it, as illustrated in the
right panel of Fig.\ref{fig:conf}. This suggests to split the
spin variable into the sum of two independent
components
\be
s_i = m_{\alpha(i)}+ \psi_i, 
\label{split.1}
\ee
where $\alpha(i)=\pm$ is the label of the
domain to which the site $i$ belongs and $\psi_i=s_i - m_{\alpha(i)}$ is, as before, 
the thermal
fluctuation variable. The significant difference with respect to the previous
case is that now $\psi_i$ and $m_{\alpha(i)}$ are both dynamical variables,
since the fluctuations of  the latter one are not due to the mixing of pure states, 
but to the transit
of the interface through the site $i$, which means that the dynamical rule updates 
both $\psi_i$ and $m_{\alpha(i)}$.
Using the independence of these variables and 
the vanishing of averages $\langle s_i \rangle_{\rm eq} = 
\langle m_{\alpha(i)} \rangle_{\rm eq} =\langle \psi_i \rangle_{\rm eq} =0$,
the correlation function can be written as the sum of two contributions
\be
C_{\rm eq}^{(a)}(r,\epsilon,L^{-1}) =  G_{\rm eq}(r,\epsilon)
+ D_{\rm eq}(r,\epsilon,L^{-1}),
\label{mix.6}
\ee
which have quite different properties.
The first one, which is the same as in Eq.~(\ref{mix.5}), is short ranged.
The $L$ dependence has been neglected, because we may always assume 
that the conditions for $\xi_\psi \ll L$ are realized.
The second one, which contains the correlations of the
background variables $m_{\alpha(i)}$, i.e.
\be
D_{\rm eq}(r,\epsilon,L^{-1}) =  
\frac{1}{V} \sum_i  \langle m_{\alpha(i)} m_{\alpha(i+r)} \rangle_{\rm eq},
\label{mix.7}
\ee
has been studied numerically in I and scales as
\be
D_{\rm eq}(r,\epsilon,L^{-1}) = \frac{1}{r^a}Y(\epsilon,r/L), \quad \text{where}
\quad Y(\epsilon,x) = m^2_\alpha (1-x).
\label{mix.8}
\ee
We have retained the power-law prefactor $r^{-a}$ in front, even though now
$a=0$ because the correlated clusters of the background variables are compact, 
in order to emphasize the 
similarity with Eq.~(\ref{sat.1}) and to render it evident by inspection
that the correlation length $\xi_m$ of these variables
coincides with $L$. Notice that $\epsilon$ does not enter the scaling function but
only its amplitude through $m^2_\alpha$.

From the divergence of $\xi_m$ in the thermodynamic limit, there
follows that the whole segment on the $\epsilon$ axis with $\epsilon < 0$
is a locus of critical points, as anticipated above. The corresponding critical
properties can be extracted by 
using $L^{-1}$ as the parameter of approach to criticality. 
It should be clear that this is bulk criticality, in no way related to
the properties of the interface, to which the attention of previous studies of the 
APBC model was primarily directed.
In I we have shown that 
the exponents satisfy the relations $\dot{\beta}/\dot{\nu} = 0$ and 
$\dot{\gamma}/\dot{\nu}=d$, where the dots identify the exponents with respect to $L^{-1}$,
e.g. from $\xi_m \sim  L$,  
follows $\dot{\nu} = 1$. This implies $\dot{\beta} = 0$ and 
$\dot{\gamma} = d$. Hence, the
hyperscaling relation $2\dot{\beta} + \dot{\gamma} = \dot{\nu}d$ is satisfied,
suggesting that the upper critical dimensionality might diverge.
So, if now we take the thermodynamic limit,
from Eqs.~(\ref{mix.6}) and~(\ref{mix.8}) we get
\be
C_{\rm eq}^{(a)}(r,\epsilon) =  G_{\rm eq}(r,\epsilon)
+ \frac{m^2_\alpha}{r^a},
\label{mix.6bis}
\ee
and, consequently, the susceptibility of the background variables
$\chi^{(a)}_m$ diverges like 
\be
\chi^{(a)}_m(\epsilon,L^{-1}) \sim  L^d,
\label{mix.9bis}
\ee
in agreement with Eq.~(\ref{susc.1}), the $m$-CK
clusters being compact.
The strong magnetization fluctuations, 
implied by the divergence of the susceptibility, are indeed exhibited by the
distribution $P^{(a)}(m)$ which, instead of being double peaked like
in Eq.~(\ref{mix.60}), has been shown in I to be uniform over the interval
$[m_-,m_+]$. The qualitative difference between
$P^{(p)}(m)$ and $P^{(a)}(m)$ is the same previously analyzed
in the $1d$ case and schematically represented in Fig.\ref{fig:1}.
The uniformity of $P^{(a)}(m)$ is the distinctive feature
which highlights the difference between condensation of fluctuations and the
usual ordering transition associated to the double-peak structure of Eq.~(\ref{mix.60}).

\subsection{Relaxation Dynamics}

\begin{figure}[ht]
\centering
\includegraphics[width=10cm]{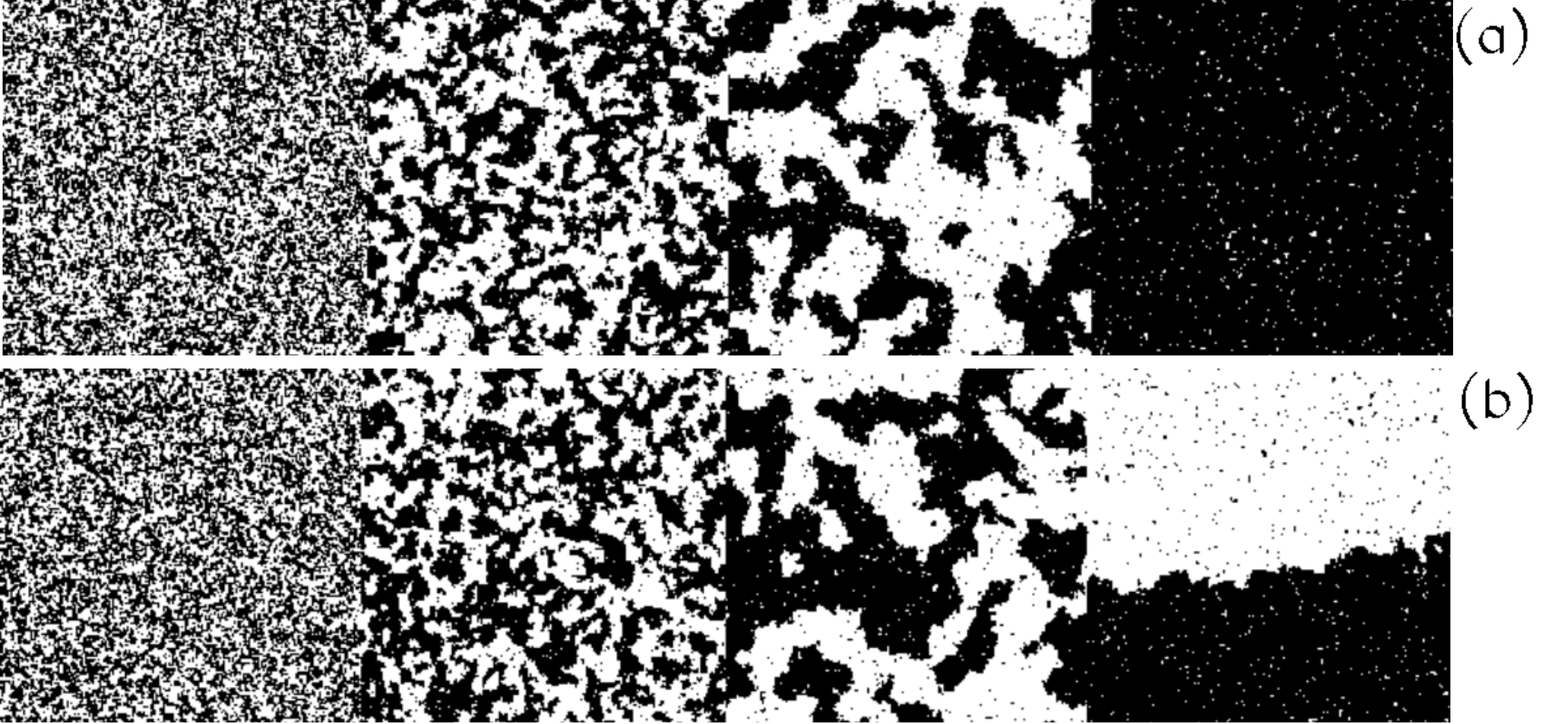}



	\caption{The two panels show the sequence of snapshots taken at
	$t=1, 10, 100, 10^5$ with PBC (a) and APBC (b), after a quench
	at $T_F=1.8$, with $L=256$. Time increases from left to right. 
	The first three configurations belong to the coarsening regime
	and show independence from BC. The last pair of configurations is 
	morphologically similar to those in Fig.~\ref{fig:conf}  and shows that at
	$t=10^5$ the system has equilibrated.}
\label{fig:conf1}
\end{figure}

When the relaxation of the infinite system is studied, by taking first
the thermodynamic limit, the dependence on BC 
is expected to disappear,
because at any finite time the correlation length is limited by $R$. This is 
confirmed by the snapshots of the typical configurations (see Fig.\ref{fig:conf1})
taken after the quench to $T_F/T_c = 0.79$. 
The top panel depicts the PBC case and the
bottom panel the APBC one. In each panel time increases from left to right.
The first three snapshots, taken at $t=1,10,100$, display the
self-similar morphology characteristic of coarsening domains, which does not show
to be affected by the
type of the imposed BC, because $R \ll L$. The BC influence is evident, instead, in the fourth snapshot taken
at $t=10^5$, when $R \gtrsim L$ and the system has equilibrated.

The configurations morphology, with large compact growing domains containing 
in their interior small patches of thermal fluctuations,
suggests to generalize
to the off-equilibrium regime the split of variables~(\ref{split.1}) by
$s_i(t) = m_{\alpha(i,t)}+ \psi_i(t)$, where
$\alpha(i,t)$ is the label of the domain to which the site $i$ belongs at
the time $t$. Then, as in Eq.~(\ref{mix.6}), the correlation
function separates into the sum of two contributions
\be
C(r,\epsilon,t^{-1}) =  G_{\rm eq}(r,\epsilon)
+ D(r;\epsilon,t^{-1}),
\label{gen.19}
\ee
where the first one is BC-independent, time-independent
and identical to the analogous term appearing in Eqs.~(\ref{mix.5}) and~(\ref{mix.6}),
because thermal fluctuations equilibrate quickly.
The second contribution contains the correlations of the background
variables and obeys scaling in the form
\be
D(r,\epsilon,t^{-1}) = \frac{m^2_\alpha}{r^a} F(r/R),
\label{gen.20}
\ee
where $a=0$, due to the compactness of domains, the growth law $R(t) = t^{1/z}$ 
is the same of Eq.~(\ref{anml.3bis})    
with $z=2$ and the $\epsilon$ dependence has been factorized in the amplitude $m^2_\alpha$. 
Comparing with Eq.~(\ref{sat.3}), we see that the same behavior as in the quench to $T_c$
is obtained, apart for the change of the exponents $z$ and $a$, and for
the specific forms of the functions $F_c(x)$ and $F(x)$.

\begin{figure}[ht]
\centering
\includegraphics[width=10cm]{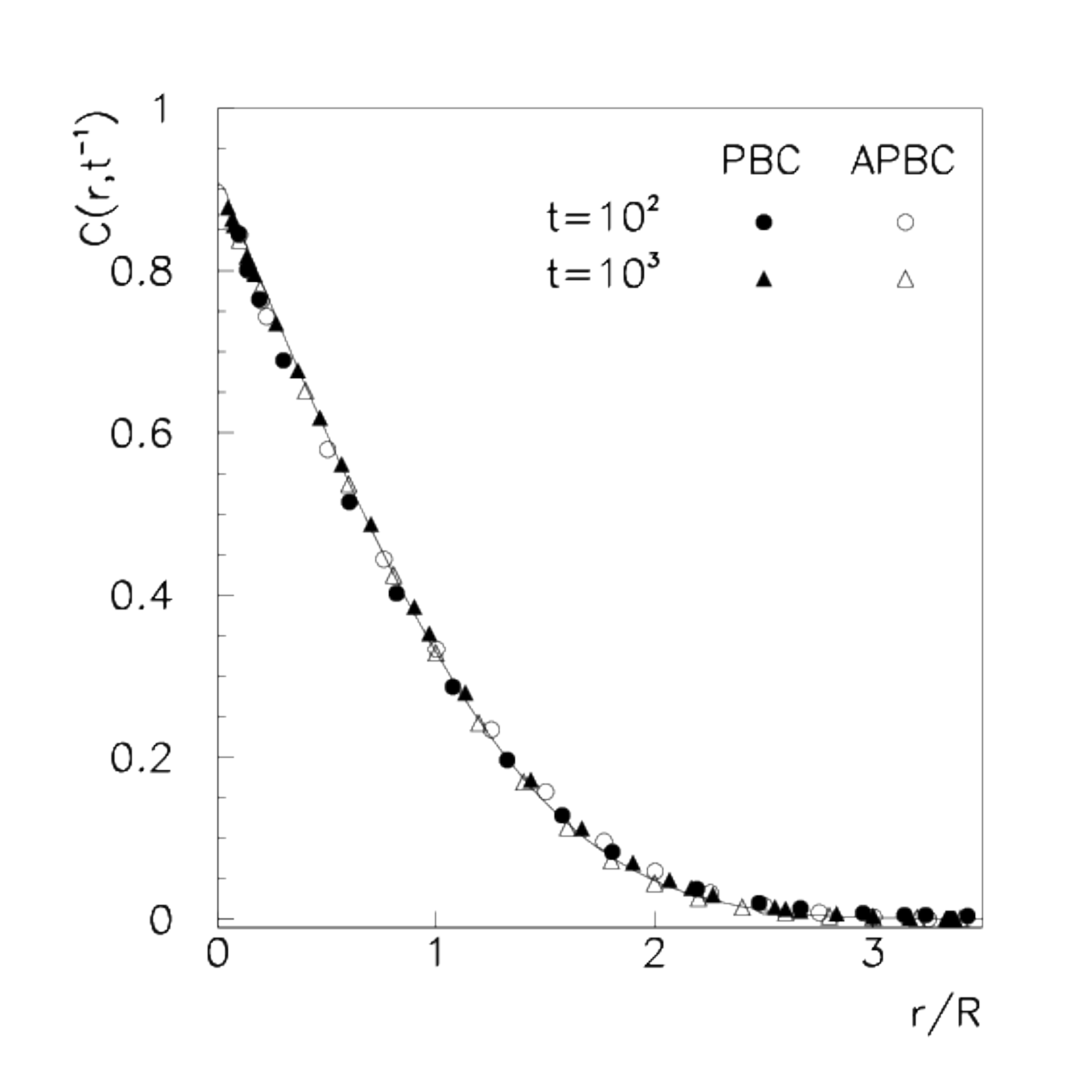}
\caption{Collapse on the master curve of Eq.~(\ref{quench.1}) of the data for
$C(r,t^{-1},L^{-1})$ in the time regime $R \ll L$, system size $L=256$. 
PBC (black symbols) and  APBC (empty symbols).
The continuous line is the plot of the Ohta-Jasnow-Kawasaki function
defined in Eq.~(\ref{OJK}).}
\label{fig:scaledCbis}
\end{figure}

The above statements are substantiated by the plot in Fig.\ref{fig:scaledCbis}
of the numerical data
for the equal-time correlation function, generated for a quench to
$T_F/T_c=0.666$, which corresponds to $\epsilon = -0.9$, with $L=256$ and with 
both PBC and APBC. The APBC data have been 
circularly averaged to smooth out the anisotropy induced by the cylindrical BC. 
The good collapse of the data, in the time
regime such that $R \ll L$, shows that for the chosen value of $T_F$ 
the thermal fluctuations contribution is negligible.
Moreover, the master curve $F(x)$ compares well with the 
Otha-Jasnow-Kawasaki~\cite{OJK} approximate result
\be
F(x) = \left (\frac{2}{\pi} \right ) \arcsin (\gamma), \quad \gamma = \exp(-x^2/b),
\label{OJK}
\ee
where $b$ is a constant, as it is demonstrated by Fig.\ref{fig:scaledCbis}. 
Therefore, the relaxation to below $T_c$ is not qualitatively different from the one
to $T_c$. Both are coarsening processes and both do not depend on the
imposed BC. Differences between the two are in the quantitative details,
like the values of the $z$ exponent, the dimensionality of correlated clusters 
and the shape of the scaling functions.
The implication is 
that also in the quench to below $T_c$ the system tends toward a 
critical state, because the time-dependent correlation length $R$ diverges,
eventually yielding the time-asymptotic critical correlation function
\be
C^*(r,\epsilon)= G_{\rm eq}(r,\epsilon) + \frac{m^2_\alpha}{r^a}.
\label{dyna.5}
\ee
It is then evident, according to the discussion made at the end
of subsection~\ref{APBCeq}, that this asymptotic form, which
we emphasize once more is the same for both choices of BC, 
matches $C^{(a)}_{\rm eq}(r;\epsilon)$ but not $C^{(p)}_{\rm eq}(r;\epsilon)$.
Finally, recalling that $a=0$, it is straightforward to see from Eq.~(\ref{gen.20})
that the background susceptibility scales like
\be
\chi (\epsilon,t^{-1}) \sim  R^d,
\label{dyna.6}
\ee
in agreement with the result~(\ref{mix.9bis}) for $\chi^{(a)}_m(\epsilon,L^{-1})$.

In conclusion, in the PBC case, as anticipated in the Introduction,
the asymptotic state and the equilibrium one are remote one from the other, and the system 
may be
regarded as remaining strongly out of equilibrium, because in the former one there are
long-range correlations, which are absent in the second one.
In the APBC case, instead, both the asymptotic and the equilibrium state are
critical and with the same universal properties, hence the system equilibrates
although with an infinite equilibration time, just as in the quench to $T_c$.

\subsection{Summary}

So far we have shown that when the Ising model is quenched in the
two-phase region, i.e. to below $T_c$ for $d=2$ and to $T_F=0$ for $d=1$, the APBC system equilibrates and the PBC one remains off equilibrium.
The basic elements of the mechanism underlying this phenomenology
are as follows:

\begin{enumerate}

\item To different BC, in principle, there correspond different statistical ensembles.

\item These ensembles become {\it equivalent} when the limits are taken according 
to the sequence: $L^{-1} \to 0$ first and then $t^{-1} \to 0$, for all temperatures $T_F$.

\item Instead, the ensembles {\it may} become {\it non equivalent}, depending on BC,
when the limits are taken in the reverse sequence: $t^{-1} \to 0$ first 
and then $L^{-1} \to 0$ with $T_F < T_c$.

\item 
Equivalence fails with PBC because of ergodicity breaking, and holds
with APBC since ergodicity is preserved.

\item Ergodicity breaking induces spontaneous symmetry breaking, which
makes correlations short-ranged.

\item Instead, when ergodicity holds an unusual type of 
criticality sets in, with long-range correlations and  {\it compact} correlated domains.

\end{enumerate}

In the next section we shall show that this is not just a peculiarity of
the Ising model, but that it is a more general phenomenon, since it 
takes place with the same characteristics also in the quench to below the critical point
of the spherical model, without invoking the imposition of different types of BC. 
In fact, two different ensembles arise not from the choice of BC, 
which is taken to be the standard PBC
one, but from enforcing the spherical constraint
either sharply or smoothly. These ensembles
turn out to be equivalent or non equivalent, just as in the Ising case, depending on
the order of the $L^{-1} \to 0$ and $t^{-1} \to 0$ limits.

\section{Spherical models}
\label{SM}

\subsection{Equilibrium}

Let us briefly recall what the spherical model is about starting from
equilibrium, which means that the $t^{-1} \to 0$ limit has been taken beforehand.
Consider a classical paramagnet in the volume $V=L^d$ and
with the energy function~\cite{Ma}
\be
{\cal H}(\boldsymbol{\varphi}) = \int_V d\vec r \, \varphi(\vec r) \left
(-\frac{1}{2}\nabla^2 \right )\varphi(\vec r),
\label{gauss.1}
\ee
where $\boldsymbol{\varphi}$ stands for a configuration  of the local, continuous
and unbounded spin variable $\varphi(\vec r)$. PBC are understood throughout.
Due to its bilinear character, the above Hamiltonian can be diagonalized by Fourier transform
\be
{\cal H} = \frac{1}{2V}\sum_{\vec k} k^2|\varphi_{\vec k}|^2.
 \label{gauss.1bis}
\ee
In the spherical model (SM) of Berlin and Kac~\cite{BK} a coupling among the modes is
induced by the imposition of an overall sharp constraint on
the square magnetization density
\be
\mathit{s}(\boldsymbol{\varphi}) = \frac{1}{V} \int_V d \vec r \,
\varphi^2(\vec r) = \frac{1}{V^2}\sum_{\vec k} |\varphi_{\vec k}|^2 = 1.
\label{gauss.2}
\ee
Then, in thermal equilibrium the statistical ensemble is given by
\be
P_\textrm{SM}(\boldsymbol{\varphi}) = \frac{1}{Z_\textrm{SM}}
e^{-\beta {\cal H}(\boldsymbol{\varphi})}
\, \delta \left (\mathit{s}(\boldsymbol{\varphi})-1 \right ),
\label{Gauss.4}
\ee
where $Z_\textrm{SM}$ is the partition function. 
A variant of the model, called
the mean-spherical model (MSM)~\cite{LW,KT}, is obtained by imposing the constraint in the mean: An exponential bias is introduced in place of the $\delta$ function
\be
P_\textrm{MSM}(\boldsymbol{\varphi}) = \frac{1}{Z_\textrm{MSM}}e^{-\beta [{\cal H}(\boldsymbol{\varphi}) +\frac{\kappa}{2} {\cal S}(\boldsymbol{\varphi})]},
\label{Gauss.3}
\ee
where ${\cal S}(\boldsymbol{\varphi}) = V\mathit{s}(\boldsymbol{\varphi})$ and the parameter $\kappa$ must be so adjusted to satisfy the requirement
\be
\langle \mathit{s}(\boldsymbol{\varphi}) \rangle_\textrm{MSM} = 1.
\label{msph.1}
\ee
Although it is the common usage to refer to these  as models, it should be clear from Eqs. (\ref{Gauss.4}) and (\ref{Gauss.3}) that we are dealing with two conjugate ensembles, distinguished by conserving or letting to fluctuate the density $\mathit{s}$.

In both models there exists a phase transition at the same critical
temperature $T_c$, above which they are equivalent and below which they are not,
which means that the nature of the low temperature phase is different. 
It is worth, here, to go in some detail~\cite{CSZ} because the point is quite illuminating
on the equivalence or lack-of issue. 
Let us separate in $\mathit{s}$ the excitations from the ground-state 
contribution
\be
\mathit{s} = \mathit{s}_0 +  \mathit{s}^*, \quad \text{with} \quad 
\mathit{s}_0 = \frac{1}{V^2}\varphi_0^2,  \quad
\quad \mathit{s}^*= \frac{1}{V^2} \sum_{\vec k \neq 0} |\varphi_{\vec k}|^2.
\label{s.1}
\ee
Then, taking the average in either ensemble, from the spherical
constraint follows the sum rule
\be
\langle  \mathit{s}_0 \rangle +  \langle \mathit{s}^* \rangle = 1,
\label{s.2}
\ee
which must be satisfied at all temperatures and it is the motor of the transition. 
In fact, in the thermodynamic limit the excitations contribution is superiorly
bounded~\cite{CSZ} by
\be
\langle \mathit{s}^* \rangle \leq TB,
\label{s.3}
\ee
where $B$ is a dimensionality-dependent positive constant, which is finite
for $d > 2$ and diverges at $d=2$. Therefore, by enforcing the
constraint~(\ref{s.2}) there remains defined
the critical temperature 
\be
T_c = 1/B,
\label{s.4}
\ee
above which the sum rule~(\ref{s.2}) is saturated without any contribution from
$\langle  \mathit{s}_0 \rangle$, while below there must necessarily be a finite contribution
from the ground state, yielding
\be
\langle  \mathit{s}_0 \rangle  = \left \{ \begin{array}{ll}
         0, \;\; $for$, \;\; T \geq T_c,\\
         1 - T/T_c,\;\; $for$, \;\; T < T_c.
        \end{array}
        \right .
        \label{s.5}
        \ee
Rewriting $\mathit{s}_0 = \psi_0^2$, where $\psi_0$ is the density $\frac{1}{V}\varphi_0$,
the question is how can there arise a finite contribution to $\langle  \mathit{s}_0 \rangle$
from this single degree of freedom and here is precisely where
the two models differ. In the SM the sharp 
version~(\ref{gauss.2}) of the constraint  introduces enough nonlinearity for the
transition to take place by {\it ordering}.
This means that ergodicity breaks down inducing the spontaneous breaking of
the $\mathbb{Z}_2$ symmetry. Then, exactly like in the Ising model with PBC,
the probability
distribution of the magnetization density, that is of $\psi_0$, results from the mixture
of the two pure ferromagnetic states
\be
P_\textrm{SM}(\psi_0) = 
\frac{1}{2}[\delta (\psi_0 - m_-) + 
         \delta (\psi_0 - m_+)],
\label{ensmbl.010}
        \ee
where $m_\pm = \pm \sqrt{1 - T/T_c}$ is the spontaneous magnetization. 
Thus, in this case $\langle  \mathit{s}_0 \rangle_{SM}$ stands for
the square of the spontaneous magnetization $m_\pm^2$.
     Instead, in the MSM ordering cannot take place, because the soft 
     version~(\ref{msph.1}) of the constraint leaves the statistics Gaussian. 
     Neither ergodicity nor symmetry break down, as in the Ising APBC case. Then, below $T_c$,
     the only mean to build up the finite value of $\langle  \mathit{s}_0 \rangle_{MSM}$
     needed to saturate the sum rule is by growing the fluctuations 
     of $\psi_0$ through the spread out probability distribution given
     by  
\be
P_\textrm{MSM}(\psi_0) =
         \frac{e^{-\frac{\psi_0^2}{2(1-T/T_c)}}}{\sqrt{2\pi (1 - T/T_c)}}.
\label{ensmbl.001}
\ee
Therefore, now $\langle  \mathit{s}_0 \rangle_{MSM}$
     stands for the macroscopic 
     variance of $\psi_0$. Elsewhere~\cite{EPL,CCZ,Zannetti,Merhav,Marsili}, this type of transition, characterized by the fluctuations of an extensive quantity condensing into one microscopic component, has been referred to as {\it condensation of fluctuations}.

\begin{figure}[!tb]
\centering
\includegraphics[width=5cm]{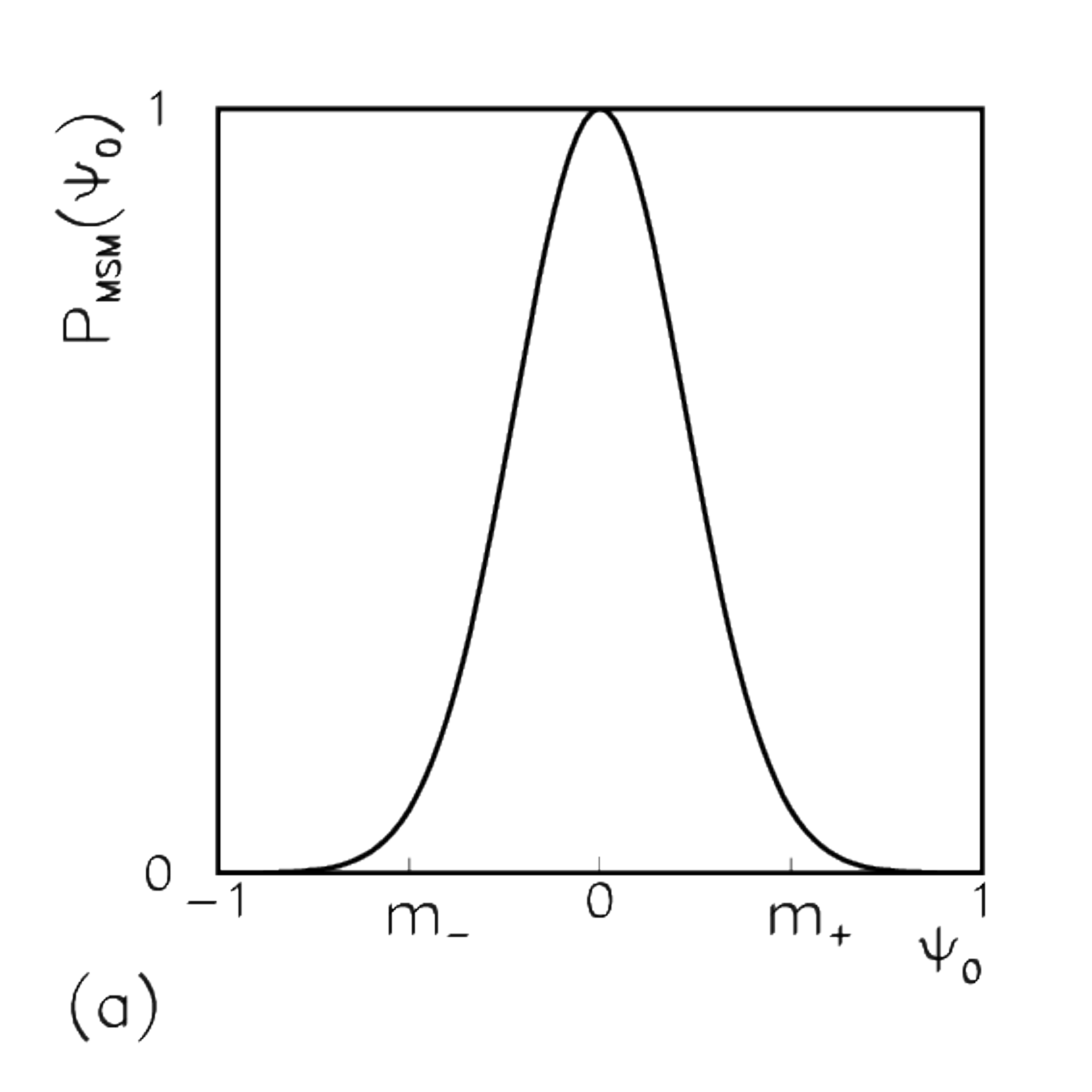} \hspace{0cm}
\includegraphics[width=5cm]{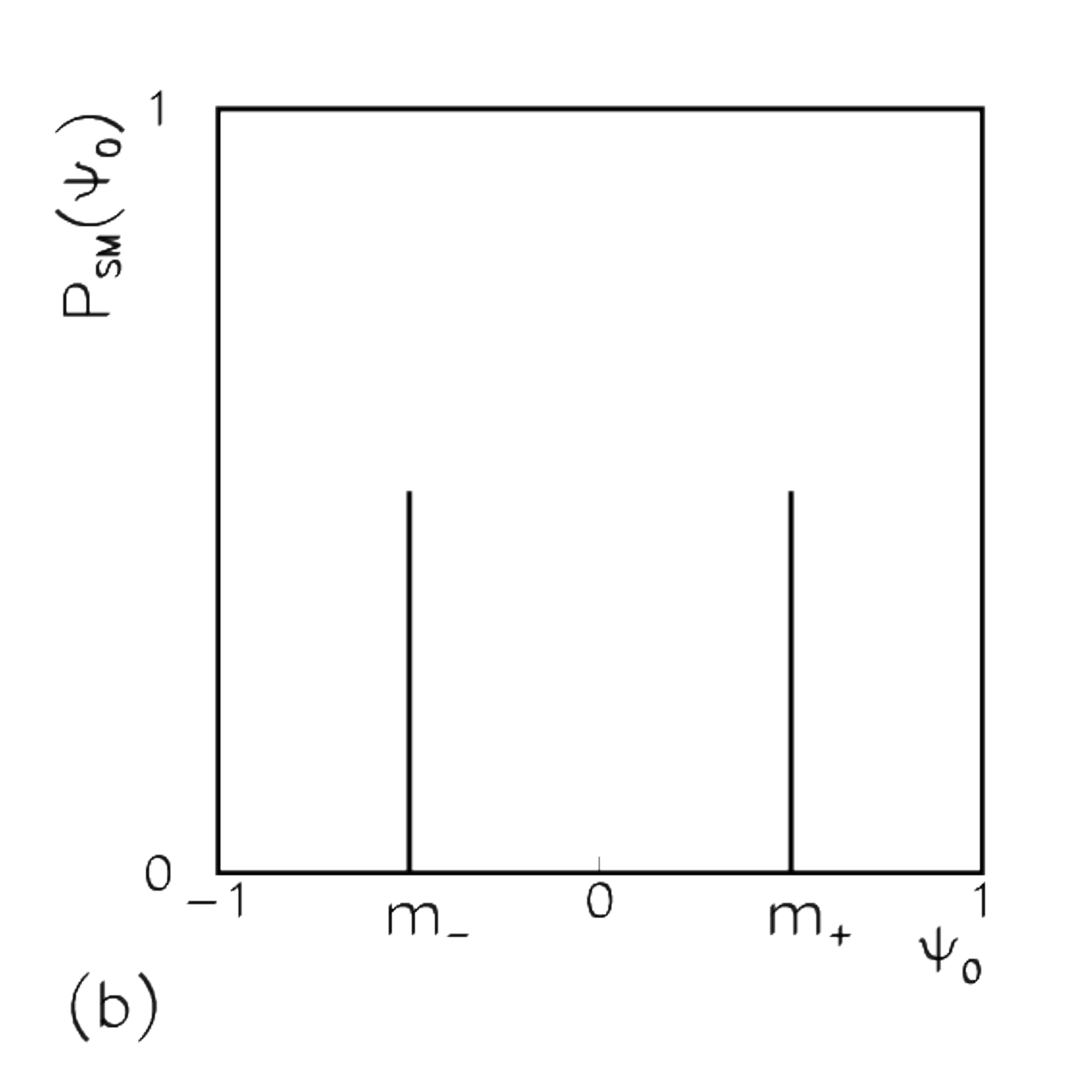}
	\caption{Magnetization distribution in the MSM model (a) and in the SM model (b),
  for $T < T_c$. The spikes in the right panel stand for $\delta$ functions.}
\label{fig1}
\end{figure}
Comparing Figs.~\ref{fig1} and~\ref{fig:1}, it is evident that the distributions are the same in the two cases
where ergodicity breaks down, that is in the Ising model with PBC and in the SM.
In the other two cases, Ising with APBC and MSM, the distributions are not 
superimposable but show the same physical phenomenon: ergodicity is preserved
by developing macroscopic fluctuations of the magnetization, which remain finite
in the thermodynamic limit and reveal
the critical nature of the low temperature phase. In fact, in the MSM the structure
factor, i.e. the Fourier transform of the correlation function, is given by~\cite{CCZ}
\be
C_\textrm{MSM}(\vec k) = \frac{T}{k^2} + m_\pm^2 \delta(\vec k).
\label{str.1}
\ee
The two terms appearing above are the anolouges
in Fourier space of those entering $C_{\rm eq}^{(a)}(r;\epsilon)$ 
in Eq.~(\ref{mix.6bis}), with the correspondences
\be
G_{\rm eq}(r;\epsilon) \longleftrightarrow \frac{T}{k^2}, \quad
\frac{m^2_\alpha}{r^a} \longleftrightarrow m_\pm^2 \delta(\vec k).
\label{str.2}
\ee
Notice that, as it is well known, the thermal fluctuations contribution in the MSM
is massless, i.e. is critical, at all temperatures below $T_c$. For simplicity, let
us set $T=0$ in order to get rid on this contribution and to focus on
the interesting one, which is the $\delta$-function term (Bragg peak).
We emphasize that this
is the Fourier transform of the background {\it critical} contribution with
compact correlated clusters, just as the corresponding term in the Ising APBC case.

Finally, we point out that the $d=2$ case is analogous to Ising with $d=1$, 
because $T_c$ vanishes.
However, for brevity, we shall not elaborate on this case here.

\subsection{Dynamics}

Let us next consider the relaxation dynamics in the quench to $T_F=0$. In Ref.~\cite{Fusco} it was shown that,
when the thermodynamic limit is taken first, the two models are equivalent
at all times. Then, it is an exact result that the dynamical structure factor 
both for the SM and MSM is given by
\be
C(\vec k,t) = \Delta \left (1 +\frac{2r^2_0}{R^2} \right )^{d/2} R^d \, e^{-(kR)^2},
\label{str.3}
\ee
where $C(\vec k,0)=\Delta$ is the spatially uncorrelated initial condition at $T_I=\infty$,
$R = \sqrt{2t}$ is the growth law for nonconserved
dynamics and $r_0 = 1/(\sqrt{2}\Lambda)$ is the microscopic length related
to the momentum cutoff $\Lambda$, which is imposed exponentially
when integrating over $\vec k$ and is responsible for the corrections to
scaling in the early regime.
From the normalization condition at $t=0$
\be
\int \frac{d^d k}{(2\pi)^d} \, C(\vec k,0) e^{-k^2/\Lambda^2} = 1,
\label{str.4}
\ee
there follows $\Delta = (2\sqrt{\pi})^d$. Inserting this into Eq.~(\ref{str.3}), with
little algebra one can verify that indeed the normalization is satisfied at all
times. Then, since the peak grows like $C(0,t) \sim R^d$, one can conclude
that the asymptotic structure factor is the $\delta$ function
\be
\lim_{t \to \infty} C(\vec k,t) = C^*(\vec k) = \delta(\vec k),
\label{str.5}
\ee
which matches the equilibrium Bragg peak~(\ref{str.1}) in the MSM. Since
the growth of $R$ implies that the asymptotic state is critical and with compact
correlated clusters, in the quench to below $T_c$ the MSM approaches
equilibrium arbitrarily close, while the SM, which is not critical in the equilibrium 
state, remains permanently out of equilibrium. 

In conclusion, going through all the items listed at the end of the previous
section, one can check that perfect correspondence between 
Ising-PBC and SM on one side and Ising-APBC and MSM on the other
is established.

\section{Concluding Remarks}
\label{CR}

In this paper we have addressed a problem which is of basic interest in the
physics of slowly relaxing systems. Since slow relaxation means that
equilibrium is not reached in the observable time scale, relevant
questions are whether a criterion for equilibration, or for lack of, can be established
and, if so, whether the nature of the equilibrium state
can be inferred from the available dynamical information. Although the task of giving
general answers to these questions is of formidable difficulty, 
we have shown that, at least in the restricted realm of phase-ordering systems,
it is possible to arrive at some definite conclusions.

By analyzing the relaxation of the Ising model after temperature
quenches, we have found that
the system does or does not equilibrate, depending on whether the dynamics at
the final temperature of the quench is ergodic or not.
This has been established by investigating the dependence
of the spin-spin correlation function upon the order of the large-time and 
thermodynamic limits, when different BC are imposed. The findings are that the
APBC system equilibrates in all conditions, because the dynamics are ergodic at
all temperatures, while the PBC system does not equilibrate for $T_F < T_c$,
because that's where ergodicity does not hold.
These statements are strengthened and corroborated by 
exact analytical results from
the quench of the spherical and mean spherical model, which reproduce
very closely, although in a quite different context, the picture just outlined. 
We may then answer
the first question asked at the beginning of the section by saying that
it might take an infinite time to equilibrate, but nonetheless
the system can get arbitrarily close to equilibrium if the dynamics are ergodic.
Instead, if ergodicity is broken, and if the initial state is symmetric,
the system does not get close to equilibrium, no matter how long is let to 
relax.

For what concerns the second question, the answer is that, yes, once it is
established that the system approaches equilibrium, then the 
nature of the equilibrium state can be inferred from the dynamical information.
Consider first the quench to $T_c$, in which case the time-dependent
correlation function obeys the scaling form~(\ref{sat.3}), while in the equilibrium
state it decays according to the pure power law~(\ref{sat.2}). It is then evident that
the latter result can be reconstructed from the short distance behaviour, 
i.e. for $r \ll R$, at times finite but large enough to detect a clean scaling behaviour.
The same procedure applies also in the case of the quench to below $T_c$ with
APBC, where the background component of the correlation function is given by
Eq.~(\ref{gen.20}). Then, again the short distance approximation, which in this case
is a constant term since $a=0$, reproduces correctly the form of the equilibrium 
critical correlation function.

Finally, let us comment on the nature of the line of critical points on the
$\epsilon < 0$ segment in the Ising APBC case. 
According to the view put forward in this paper,
$t^{-1}$ is just another relevant parameter measuring
the distance from criticality, on the same footing with $\epsilon$ and $L^{-1}$, so that
these critical points control both statics and dynamics. In subsection \ref{APBCeq}
we have pointed out
that the static critical exponents, defined with respect to $L^{-1}$, 
satisfy the hyperscaling relation 
$2\dot{\beta} + \dot{\gamma} = \dot{\nu }d$ for all $d$, suggesting that 
the upper critical dimension is at $d=\infty$. It is then interesting to note 
the concomitance with the fact that the 
Otha-Jasnow-Kawasaki approximate theory, which accounts well for the time
dependent correlation function as shown in Fig.\ref{fig:scaledCbis},
becomes exact in the $d \to \infty$ limit~\cite{Bray}.

\begin{acknowledgements}
A.F. acknowledges financial support of the MIUR PRIN 2017WZFTZP "Stochastic forecasting in complex systems".
\end{acknowledgements}




\begin{thebibliography}{99}


\bibitem{Palmer}
R. G. Palmer, Adv. in Phys. {\bf 31}, 669 (1982).

\bibitem{BCKM}
J. P. Bouchaud, L. F. Cugliandolo, J. Kurchan and M. Mezard, Out of equilibrium 
dynamics in spin glasses and other glassy systems, in 
{\it Spin Glasses and Random Fields}, edited by A. P. Young (World Scientific, Singapore, 1997); arXiv:cond-mat/9702070.

\bibitem{Bray}
A. J. Bray, Adv. Phys. {\bf 43}, 357 (1994).

\bibitem{Puri}
S. Puri, in {\it Kinetics of Phase Transitions}, edited by S. Puri and V. Wadahawan (CRC Press, Boca Raton, FL, 2009).

\bibitem{Jo}
M. Zannetti, in {\it Kinetics of Phase Transitions}, edited by S. Puri and V. Wadahawan (CRC Press, Boca Raton, FL, 2009).

\bibitem{Henkel}
M. Henkel and M. Pleimling, {\it Non-Equilibrium Phase Transitions}, Vol. 2,
(Springer, Dordrecht, 2010).

\bibitem{FCZ}
A. Fierro, A. Coniglio and M. Zannetti, Phys. Rev. E {\bf 99}, 042122 (2019).

\bibitem{Gallavotti}
G. Gallavotti, Riv. Nuovo Cimento {\bf 2}, 133 (1972);
{\it Statistical Mechanics A Short Treatise}, Springer-Verlag Berlin Heidelberg 1999. 

\bibitem{Antal}
T. Antal, M. Droz and Z. R\'acz, J. Phys. A: Math. Gen. {\bf 37}, 1465 (2004). 

\bibitem{Janssen}
H. K. Janssen, B. Schaub and B. Schmittman, Z. Phys. B {\bf 73}, 539 (1989).

\bibitem{CK}
A. Coniglio and W. Klein, J. Phys. A {\bf 13}, 2775 (1980).


\bibitem{FK}
W. P. Kasteleyn and C. M. Fortuin, J. Phys. Soc. Japan Suppl. {\bf 26}, 11 (1969).

\bibitem{CF}
A. Coniglio and A. Fierro, Correlated Percolation, in {\it Encyclopedia of Complexity and Systems Science}, Part 3, edited by R. A. Meyers (Springer-Verlag, New York, 2009),
pp. 1596-1615; arXiv:1609.04160.

\bibitem{Stanley}
H. E. Stanley, {\it Introduction to Phase Transitions and Critical Phenomena},
Oxford Science Publications.

\bibitem{Goldenfeld}
N. Goldenfeld, {\it Lectures on Phase Transitions and the Renormalization Group},
Addison-Wesley (1972).

\bibitem{1d}
A. J. Bray, J. Phys. A {\bf 22}, L67 (1990);  J. G. Amar and F. Family, Phys. Rev. A
{\bf 41}, 3258 (1990); B. Derrida, C. Godr\`eche and I. Yekutieli Phys. Rev. A, {\bf 44}, 6241 (1991).

\bibitem{Nightingale}
M. P. Nightingale and H. W. J. Bl\"{o}te, Phys. Rev. Lett. {\bf 76}, 4548 (1996).

\bibitem{Das}
S. K. Das, S. Roy, S. Majumder and S. Ahmad, Europhys. Lett. {\bf 97}, 66006 (2012).

\bibitem{Binder}
K. Binder, Z. Phys. B {\bf 43}, 119 (1981).

\bibitem{Bruce}
A. D. Bruce, J. Phys. C: Solid State Phys. {\bf 14}, 3667 (1981);
{\it ibidem} {\bf 18}, L873 (1985).










\bibitem{OJK}
T. Ohta, D. Jasnow and K. Kawasaki, Phys. Rev. Lett. {\bf 49}, 1223 (1982).














\bibitem{Ma}
S. K. Ma, {\it Modern Theory of Critical Phenomena}, W A Benjamin (1976);
D. J. Amit and V. Mart\'in-Mayor, {\it Field Theory, the Renormalization Group, and Critical Phenomena}, World Scientific Singapore (2005).

\bibitem{BK}
T. H. Berlin and M. Kac, Phys. Rev. {\bf 86}, 821 (1952).

\bibitem{LW}
H. W. Lewis and G. H. Wannier, Phys. Rev. {\bf 88}, 682 (1952) and
Phys. Rev. {\bf 90}, 1131E (1953).

\bibitem{KT}
M. Kac and C. J. Thompson, J. Math. Phys. {\bf 18}, 1650 (1977).






\bibitem{CSZ}
A. Crisanti, A. Sarracino and M. Zannetti, Phys. Rev.  R {\bf 1}, 023022 (2019).

\bibitem{EPL}
M. Zannetti, EPL {\bf 111}, 20004 (2015).

\bibitem{CCZ}
C. Castellano, F. Corberi and M. Zannetti, Phys. Rev. E {\bf 56}, 4973 (1997).

\bibitem{Zannetti}
F. Corberi, G. Gonnella, A. Piscitelli and M. Zannetti, J. Phys. A: Math. Theor. {\bf 46}, 042001 (2013);
M. Zannetti, F. Corberi and G. Gonnella, Phys. Rev. {\bf E 90}, 012143 (2014); M.Zannetti,
F. Corberi, G. Gonnella and A. Piscitelli, Commun. Theor. Phys. {\bf 62}, 555 (2014).

\bibitem{Merhav}
N. Merhav and Y. Kafri, J. Stat. Mech. P02011 (2010).

\bibitem{Marsili}
M. Filiasi, G.Livan, M. Marsili. M. Peressi, E. Vesselli and E. Zarinelli, J. Stat. Mech. P09030 (2014); 
M. Filiasi, E. Zarinelli, E. Vesselli and M. Marsili, arXiv:1309.7795v1;
L. Ferretti, M. Mamino and G. Bianconi, Phys. Rev. E {\bf 89}, 042810 (2014).

\bibitem{Fusco}
N. Fusco and M. Zannetti, Phys. Rev. E {\bf 66}, 066113 (2002).
\end{thebibliography}
\end{document}